\documentclass[12pt]{article}

\usepackage{amssymb}
\usepackage[dvips]{graphicx}

\newcommand {\beq}{\begin{equation}}
\newcommand {\eeq}{\end{equation}}
\newcommand {\beqa}{\begin{eqnarray}}
\newcommand {\eeqa}{\end{eqnarray}}
\newcommand {\n}{\nonumber \\}
\newcommand {\tr}{\mbox{tr}}

\def\pa{\partial}
\renewcommand{\theequation}{\thesection.\arabic{equation}}

\begin{document}
\setlength{\oddsidemargin}{0cm}
\setlength{\baselineskip}{7mm}

\begin{titlepage}
\renewcommand{\thefootnote}{\fnsymbol{footnote}}
\begin{normalsize}
\begin{flushright}
\begin{tabular}{l}
OU-HET 445 \\
May 2003
\end{tabular}
\end{flushright}
  \end{normalsize}

~~\\

\vspace*{0cm}
    \begin{Large}
       \begin{center}
         {Hamilton-Jacobi Method and Effective Actions \\
          of D-brane and M-brane in Supergravity}
       \end{center}
    \end{Large}
\vspace{1cm}

\begin{center}
           Matsuo S{\sc ato}\footnote
            {
e-mail address : 
machan@het.phys.sci.osaka-u.ac.jp}
           {\sc and}
           Asato T{\sc suchiya}\footnote
           {
e-mail address : tsuchiya@het.phys.sci.osaka-u.ac.jp}\\
      \vspace{1cm}
       
        {\it Department of Physics, Graduate School of  
                     Science}\\
               {\it Osaka University, Toyonaka, Osaka 560-0043, Japan}\\
    \end{center}

\hspace{5cm}

\begin{abstract}
\noindent
We show that the effective actions of D-brane and M-brane are
solutions to the Hamilton-Jacobi (H-J) equations in supergravities.
This fact means that these effective actions are on-shell actions 
in supergravities. These solutions to the H-J equations reproduce
the supergravity solutions that represent D-branes in a 
$B_2$ field, M2 branes and the M2-M5 bound states. The
effective actions in these solutions are those of a probe D-brane and
a probe M-brane. Our findings can be applied to the study of the gauge/gravity
correspondence, especially the holographic renormalization group, and a search
for new solutions of supergravity.
\end{abstract}
\vfill
\end{titlepage}
\vfil\eject

\setcounter{footnote}{0}

\section{Introduction}
\setcounter{equation}{0}
\renewcommand{\thefootnote}{\arabic{footnote}} 
%The Hamilton-Jacobi method is the most sophisticated one for solving
%equations of motion in classical mechanics.
In this paper, we show that the D-brane effective action (the Born-Infeld
action plus the Wess-Zumino action) is a solution to the Hamilton-Jacobi (H-J)
equation of type IIA(IIB) supergravity 
and that the M-brane effective action is 
a solution to the H-J equation of 11-dimensional (11-d) supergravity.
This fact means that the effective actions of D-brane and M-brane 
are on-shell actions in supergravities. We also show that these solutions
to the H-J equations reproduce 
the supergravity solutions which represent a stack 
of D-branes in a $B_2$ field,
a stack of M2-branes and a stack of the M2-M5 bound states.
In fact, we reported the case of 
D3-brane in our previous publication \cite{ST}, and in this paper
we generalize
our previous result to the cases of D$p$-branes and M-branes.

The D-brane effective action on a curved background is obtained
in principle by calculating the disk
amplitude in superstring on the background.
The disk amplitude in the open string picture is translated
into the closed string picture as the transition 
amplitude between the vacuum and the boundary state 
representing a probe D-brane. 
This transition amplitude should reduce in the $\alpha' \rightarrow 0$ limit
to an on-shell action 
in type IIA(IIB) supergravity, which is a functional
of the values of the fields on a boundary,
Therefore, the $\alpha' \rightarrow 0$ limit of the D-brane effective
action should be a solution to the H-J equation of type IIA(IIB) supergravity.
Considering the gauge invariance, we see 
that the $\alpha' \rightarrow 0$ limit of the D-brane effective action
corresponds to setting the combination of 
the gauge field plus the NS 2-form
field to be zero.
Nevertheless, a nontrivial fact we obtain is that the D-brane effective action 
itself is a solution
to the H-J equation. Probably this fact comes from
the supersymmetry, since the R-R plays a crucial role in our analysis.

The strategy of our analysis is as follows. We reduce type IIA(IIB)
supergravity on $S^{8-p}$, dropping the fermionic degrees of freedom
consistently, and obtain a $(p+2)$-dimensional gravity.
Adopting the radial coordinate
as time, we develop the canonical formalism based on the ADM decomposition
for this $(p+2)$-dimensional gravity and obtain the H-J equation originating
from the hamiltonian constraint.
We solve the H-J equation under the condition that the fields be constant
on fixed-time surfaces, and find that the D$p$-brane effective action is
a solution to the H-J equation and reproduces the supergravity solution
of a stack of D$p$-branes in a $B_2$ field. We note here that
the near-horizon limit of this supergravity solution with $p=3$
is conjectured to be dual to noncommutative Yang Mills (NCYM) \cite{SW,HI,MR}
and reduces to $AdS_5 \times S^5$ in the commutative limit, 
which is dual to ${\cal N}=4$ super Yang Mills \cite{Maldacena,GKP-Witten}.
In our formulation, the fixed-time surface whose dimension is
$p+1$ can be interpreted as the worldvolume of a probe D$p$-brane, 
and the radial time as the position of the probe D$p$-brane.
We also reduce 11-d supergravity on $S^7$ and $S^4$,
repeat the above steps, and obtain the M2 and M5 brane effective actions as
solutions to the H-J equations, respectively. We find that these solutions
to the H-J equation reproduce the supergravity solutions of a stack of
M2-branes and a stack of the M2-M5 bound states, respectively.
Furthermore, by using the $SL(2,R)$ symmetry in type IIB supergravity
and the relation of type IIA supergravity with 11-d supergravity,
we obtain solutions to the H-J equations that reproduce
the supergravity solutions representing
$(p,q)$ strings and $(p,q)$ 5-branes in type IIB supergravity and
NS 5-branes in type IIA supergravity. Note that the near-horizon limit of
supergravity backgrounds reproduced by M2-branes,
the M2-M5 bound states and NS 5-branes are conjectured to be dual to
three-dimensional ${\cal N}=8$ superconformal field theory,
(a noncommutative version of) six-dimensional ${\cal N}=(2,0)$ superconformal 
field theory and Little String
Theory, respectively \cite{Maldacena,MR,ABKS}.

As we discuss below, the fact that our solutions to the H-J equations are the
on-shell actions around the supergravity backgrounds which conjectured to
be dual to various gauge theories
motivates us to further investigate the subject in the present paper.

Indeed, our findings can be applied to the study of the gauge/gravity 
correspondence. A well-understood example of the gauge/gravity correspondence 
is the AdS/CFT correspondence \cite{Maldacena,GKP-Witten}; in particular, 
the correspondence between 
${\cal N}=4$ super Yang Mills at the conformally
invariant point and type IIB supergravity on $AdS \times S^5$.
It is relevant to investigate whether this kind of correspondence 
can be extended to
${\cal N}=4$ super Yang Mills in the Coulomb branch \cite{Maldacena,DT}
or four-dimensional less supersymmetric
(${\cal N}=0,1,2$) gauge theories \cite{lesssusy} 
or higher (lower) dimensional 
supersymmetric gauge theories \cite{Maldacena,ABKS,JKY}. 
Another relevant problem 
to be studied from the viewpoint of the gauge/gravity correspondence is 
a quantum theory of NCYM. Classical aspects of NCYM such as noncommutative
instantons \cite{NCinstanton,SW} are well-understood while little is known 
about quantum aspects.
In particular, renormalizability of NCYM has not been established 
perturbatively or non-perturbatively. Note that the authors of Ref.\cite{BHN}
verified the renormalizability of two-dimensional bosonic NCYM by
performing a numerical simulation of its lattice version. 
As we describe shortly, solving the H-J equation
and obtaining the on-shell action in supergravity
is doubly important for studying
the above issues in the gauge/gravity correspondence. In order to
study the gauge/gravity correspondence for the less supersymmetric
gauge theories, we must generalize our analysis in the present paper; we 
should reduce 
supergravities in more complicated ways and search for solutions to the
H-J equations of these reduced gravities that reproduce 
the supergraviy solutions
conjectured to be dual to the less supersymmetric gauge
theories.

One application of the H-J method in supergravity to the study of the
gauge/gravity correspondence is to compare 
the on-shell action of supergravity with the effective action of the dual
gauge theory. Suppose that supergravity on a certain background corresponds 
to a large $N$ (noncommutative) gauge theory in which 
one of the Higgs fields has a nontrivial vacuum expectation value (vev)
and the $U(N)$ gauge symmetry is spontaneously broken to $U(1) \times SU(N-1)$.
Then, if we interpret the radial time as a position of the probe D-brane,
the on-shell action of supergravity 
around this background should coincide with the effective action of 
the gauge theory via the identification of the radial time
with the vev of the Higgs field. Thus the H-J method is useful
for checking this case of the gauge/gravity correspondence. 
It is actually conjectured in Ref.\cite{BPT} that the effective action of
${\cal N}=4$ super Yang Mills in the Coulomb branch takes the form of
the D3-brane effective action in the 't Hooft limit.
It is important to perform a similar calculation of the effective action
in NCYM and to compare the result with the on-shell action obtained in
this paper. 

The other application is the study of 
the holographic renormalization group, which is also useful for establishing
the gauge/gravity correspondence. The authors of Ref.\cite{dBVV}
derived the renormalization group equation in the dual gauge theory
from the H-J equation in supergravity. In particular, they found that 
in their simple examples
the lowest term in the derivative expansion of the on-shell action 
in supergravity plays a role of the counter terms and gives the beta functions 
and the anomalous dimensions in the dual gauge theory. In this context, the
radial time is interpreted as the renormalization scale 
in the dual gauge theory.
Note that the solutions found in this paper are also the
lowest term in the derivative expansion. One can check whether supergravity
on a certain background corresponds to a large $N$ gauge theory
by comparing the beta functions and the anomalous dimensions given by
the on-shell action around the background 
with those in the gauge theory. Also, one can examine the structure of
the renormalization of NCYM through the holographic renormalization group.
Although it is not obvious whether
the lowest term in the on-shell action is sufficient in more complicated cases 
we are interested in,
our results are at least a first step to the study of the holographic 
renormalization group in these cases.

Another application of the H-J method is searching for new solutions 
and classifying the solutions in supergravity. 
Using the solution to the H-J equation obtained in the present paper,
we expect to
be able to obtain new solutions
in supergravity reduced on higher dimensional spheres under the condition
that the fields depend only on the radial coordinate. 
That is, while the supergravity solution representing
D$p$-branes in a $B_2$ field
is obtained by T-dualizing a tilted smeared D$(p-1)$-brane solution 
\cite{BMM-CP}, 
we can search supergravity solutions that cannot be obtained 
by such a T-dualization, as is discussed below. If we find a solution
to the H-J equation of supergravity
reduced in a different way, we expect to be able to obtain a different kind
of new solutions of supergravity.
Furthermore, if we find a complete solution to the H-J equation under a certain
condition, we can classify the solutions in supergravity under the condition.
Hence, our analysis in the present paper should be a first step to the 
classification of the supergravity solutions.

As is well-known, a complete solution to
the H-J equation that includes as many arbitrary constants as the number
of the degrees of freedom is a generator of the canonical transformation
that makes the new Hamiltonian vanish. If one finds a complete solution to
the H-J equation, one can represent the coordinates as functions of time for
an arbitrary initial condition. Namely, the problem can be completely solved.
Although our solution to the H-J equation
is not a complete solution, it includes several arbitrary constants. 
We can reduce the
original equations of motion in supergravity, which are the second order
differential equations, to the first order ones by using the solution, and
obtain as many conserved quantities as the number of the arbitrary constants.
We may solve the first order equations to find a new solution of supergravity,
utilizing these conserved quantities. Alternatively, we can search 
for a new solution to the
H-J equation that generalizes the present solution and includes more arbitrary
constants so that we may find a new solution more easily.

The present paper is organized as follows. In section 2, we develop
the Hamilton-Jacobi method in general constrained systems. In section 3, we
perform reductions of supergravities on higher-dimensional spheres.
In section 4, we develop
the canonical formalism for the reduced gravities obtained in section 3
to derive the H-J equations. In section 5, we find 
that the D$p$-brane effective action is a solution to the H-J equation of
the reduced gravity and reproduces the supergravity solution of a stack of 
D3-branes in a $B_2$ field. Section 6 is devoted to 
the similar calculation in the cases of M2-brane and M5-brane. 
In section 7, using the SL(2,R) symmetry in type IIB supergravity and
the relation between 11-d supergravity and type IIA supergravity,
we obtain solutions to the H-J equations that reproduce
the supergravity solutions of $(p,q)$ string and $(p,q)$ 5-brane in
type IIB supergravity and of NS 5-brane in type IIA supergravity.
Section 8 is devoted to discussion and perspective. In appendix A,
the equations of motion in supergravities are listed. In appendix B, we
write down the ansatz for the fields 
made in performing a reduction of type IIA(IIB) supergravity on $S^{8-p}$.

\vspace{1cm}

\section{H-J method in constrained systems} 
\setcounter{equation}{0}
Let us consider a classical system with $n$ degrees of freedom and 
$k$ first class constraints.
Suppose the action of the system is given in the canonical form:
\beqa
I=\int^{t}_{t_0} d\tau (p_i \dot{q}_i -H(p_1,\cdots.p_n,q_1,\cdots,q_n,\tau)
-\alpha_a f_a(p_1,\cdots.p_n,q_1,\cdots,q_n,\tau)),
\label{action}
\eeqa
where $i$ runs from 1 to $n$, $a$ runs from 1 to $k$, $H$ is the Hamiltonian
and the $\alpha_a$ are Lagrange multipliers.
By variating the action with respect to $q_i$, $p_i$ and $\alpha_a$, one 
obtains the following equations of motion:
\beqa
&&\dot{q}_i=\frac{\pa H}{\pa p_i} + \alpha_a \frac{\pa f_a}{\pa p_i}, \n
&&\dot{p}_i=-\frac{\pa H}{\pa q_i} - \alpha_a \frac{\pa f_a}{\pa q_i}, \n
&&f_a(p_1,\cdots,p_n,q_1,\cdots,q_n,\tau)=0.
\label{eom}
\eeqa
Let $q_i=\bar{q}_i(\tau)$ and $p_i=\bar{p}_i(\tau)$ be a solution to the above 
equations of motion which satisfies the boundary conditions $\bar{q}_i(t)=x_i$.
Substituting this classical solution to the action gives the on-shell action,
which can be regarded as a function of the boundary positions $x_i$ and the
final time $t$.
\beqa
S(x_1,\cdots,x_n,t)=\int^{t}_{t_0} d\tau
(\bar{p}_i \dot{\bar{q}}_i 
-H(\bar{p}_1,\cdots.\bar{p}_n,\bar{q}_1,\cdots,\bar{q}_n,\tau)).
\eeqa
The variation of $S$ with respect to $x_i$ and $t$ is given by
\beqa
&&\delta S=
(\bar{p}_i(t) \dot{\bar{q}}_i(t) -H(\bar{p}(t),\bar{q}(t),t))\delta t \n
&&\qquad\;\; +\int^{t}_{t_0} d\tau 
\left(\delta \bar{p}_i \dot{\bar{q}}_i +\bar{p}_i \delta \dot{\bar{q}}_i
-\frac{\pa H(\bar{p},\bar{q},\tau)}{\pa \bar{p}_i}\delta \bar{p}_i
-\frac{\pa H(\bar{p},\bar{q},\tau)}{\pa \bar{q}_i}\delta \bar{q}_i \right).
\label{delS1}
\eeqa
Using the equations of motion (\ref{eom}) and integrating 
the righthand side of (\ref{delS1}) partially, one obtains
\beqa
&&\delta S=
(\bar{p}_i(t) \dot{\bar{q}}_i(t) -H(\bar{p}(t),\bar{q}(t),t))\delta t
+\bar{p}_i(t)\delta \bar{q}_i(t) \n
&&\qquad\;\; +\int^{t}_{t_0} d\tau 
\alpha_a \left( 
\frac{\pa f_a(\bar{p},\bar{q},\tau)}{\pa \bar{p}_i}\delta \bar{p}_i
+\frac{\pa f_a(\bar{p},\bar{q},\tau)}{\pa \bar{q}_i}\delta \bar{q}_i \right).
\label{delS2}
\eeqa
Here the last line in (\ref{delS2}) vanishes, since
\beqa
0=f_a(\bar{p}+\delta \bar{p},\bar{q}+\delta \bar{q},\tau)
=f_a(\bar{p},\bar{q},\tau).
\nonumber
\eeqa
Noting that 
$\bar{q}_i(t+\delta t)+\delta \bar{q}_i(t+\delta t)=x_i+\delta x_i$,
one obtains $\delta \bar{q}_i(t)=\delta x_i - \dot{\bar{q}}_i(t)\delta t$.
Therefore, (\ref{delS2}) reduces to
\beqa
\delta S = -H(\bar{p}(t),x,t)\delta t +\bar{p}_i(t) \delta x_i,
\eeqa
which is equivalent to the equations
\beqa
\frac{\pa S}{\pa t}&=&-H(\bar{p}(t),x,t), \n
\frac{\pa S}{\pa x_i}&=&\bar{p}(t)_i.
\label{delSdeltanddelSdelx}
\eeqa
Finally, one obtains the equations satisfied by the on-shell action $S$,
the H-J equations:
\beqa
&&\frac{\pa S}{\pa t} 
+H\left(\frac{\pa S}{\pa x_1},\cdots,\frac{\pa S}{\pa x_n},x_1,\cdots,x_n,t 
\right)=0, \n
&&f_a\left(\frac{\pa S}{\pa x_1},\cdots,\frac{\pa S}{\pa x_n},x_1,\cdots,x_n,t 
\right)=0.
\label{fullHJ}
\eeqa

Suppose a solution to (\ref{fullHJ}) is given. Then, the canonical momenta
are represented in terms of $x_i$ by using the second equation in
(\ref{delSdeltanddelSdelx}). It follows that the equations of motion 
(\ref{eom}) reduce to a set of the first order differential equations
only for $q_i$. Thus one can simplify the problem of solving the equations
of motion.

It is well-known that the H-J equation is in general more powerful as seen
in below.
First, note that if $S$ is a solution to the above equations, 
$S+\sigma$ is also
a solution to it, where $\sigma$ is a arbitrary constant.
Let the above solution possess $l$ arbitrary constants
$\beta_1,\cdots,\beta_l$ that are not the
trivial additive constant $\sigma$. Then, the following quantities are
conserved quantities:
\beqa
\gamma_s=\frac{\pa S}{\pa \beta_s} \;\;\;\;(s=1,\cdots,l).
\label{gamma}
\eeqa
In fact, on one hand,
\beqa
\frac{d}{dt}\gamma_s
=\frac{\pa^2 S}{\pa \beta_s\pa t}+\frac{\pa^2 S}{\pa\beta_s\pa x_i}\dot{x}_i.
\label{delgamma}
\eeqa
On the other hand, from (\ref{fullHJ}) and the arbitrariness of $\beta_s$, 
we obtain
\beqa
&&0=\frac{\pa}{\pa \beta_s}\left(\frac{\pa S}{\pa t}+H\right)
=\frac{\pa^2 S}{\pa\beta_s\pa t}+\frac{\pa H}{\pa (\pa S/\pa x_i)}\frac{\pa^2 S}{\pa\beta_s\pa x_i} \n
&&0=\alpha_a \frac{\pa}{\pa\beta_s}f_a=\alpha_a \frac{\pa f_a}{\pa (\pa S/\pa x_i)}\frac{\pa^2 S}{\pa\beta_s\pa x_i}.
\label{betaderivative}
\eeqa
Summing the two equations in (\ref{betaderivative}) and using (\ref{eom}) 
and (\ref{delgamma}) gives
\beqa
\frac{d}{dt}\gamma_s=0.
\eeqa

In particular, when $l=n$, it follows that $\beta_i$ and $\gamma_i$ are
new canonical momentum and new coordinates which are obtained from the
canonical transformations generated by $S$, respectively,
and are constant with respect to
the time. One can completely solve the problem by using the second equation in 
(\ref{delSdeltanddelSdelx}) and (\ref{gamma}).

\vspace{1cm}

\section{Reductions of type IIA(IIB) and 11-d supergravities on higher
dimensional spheres}
\setcounter{equation}{0}
In this section, we perform reductions of supergravities on higher
dimensional spheres. We will regard the fixed-time surface as the worldvolume
of the $p$-brane
(D$p$-brane or M2-brane or M5-brane) later. Hence, we will 
reduce supergravity to a $(p+2)$-dimensional gravity. 

First, in order to fix the conventions, we write down the actions of
type IIA(IIB) and 11-d supergravities.
In this paper, we drop the fermionic degrees of freedom consistently.
In the following equations, 
$|K_q|^2=\frac{1}{q!} K_{M_1 \cdots M_q} K^{M_1 \cdots M_q}$ for a
$q$-form $K_q$, where the appropriate metric is used for the contractions,
and $C_{p+1}$ is the R-R $(p+1)$-form. 
The bosonic part of type IIA supergravity is given by
\beqa
&&I_{IIA}=\frac{1}{2\kappa_{10}^{\:2}} \int d^{10} X \sqrt{-G}
\left[ e^{-2\phi} \left(R_G+4\pa_M \phi \pa^M \phi
-\frac{1}{2}|H_3|^2\right) 
\right. \n
&& \left.  \qquad \qquad \qquad \qquad \qquad \qquad \;\;\;
-\frac{1}{2}|F_2|^2-\frac{1}{2}|\tilde{F}_4|^2  \right]\n
&& \qquad \;\;\; -\frac{1}{4\kappa_{10}^{\:2}} 
\int B_2 \wedge F_4 \wedge F_4,
\label{IIAaction}
\eeqa
where
\beqa
&&H_3=d B_2, \;\; F_{p+2}=d C_{p+1} \;\; (p=0,2), \n
&&\tilde{F}_4=F_4-C_1 \wedge H_3. 
\eeqa
The bosonic part of type IIB supergravity is given by
\beqa
&&I_{IIB}=\frac{1}{2\kappa_{10}^{\:2}} \int d^{10} X \sqrt{-G}
\left[ e^{-2\phi} \left(R_G+4\pa_M \phi \pa^M \phi
-\frac{1}{2}|H_3|^2\right) 
\right. \n
&& \left. \qquad \qquad \qquad \qquad \qquad \qquad \;\;\;
-\frac{1}{2}|F_1|^2-\frac{1}{2}|\tilde{F}_3|^2
-\frac{1}{4}|\tilde{F}_5|^2 \right] \n
&& \qquad \;\;\; +\frac{1}{4\kappa_{10}^{\:2}} 
\int C_4 \wedge H_3 \wedge F_3,
\label{IIBaction}
\eeqa
where
\beqa
&&H_3=d B_2, \;\; F_{p+2}=d C_{p+1} \;\; (p=-1,1,3), \n
&&\tilde{F}_3=F_3+C_0 \wedge H_3, \n
&&\tilde{F}_{5}=F_{5}+C_2 \wedge H_3.
\eeqa
One must also impose the self-duality condition
\beq
\ast \tilde{F_5}=\tilde{F_5}
\label{self-duality}
\eeq
on the equations of motion derived from the above action. 
%For completeness,
%we list all the equations of motion and the self-duality condition in
%type IIB supergravity explicitly in appendix A.
The bosonic part of 11-d supergravity is given by
\beqa
I_{11}=\frac{1}{2\kappa_{11}^{\:2}} \int d^{11}X \sqrt{-G}
\left(R_G-\frac{1}{2}|F^{(M)}_4|^2 \right) 
-\frac{1}{12\kappa_{11}^{\:2}} \int A_3 \wedge F^{(M)}_4 \wedge F^{(M)}_4,
\label{11Daction}
\eeqa
where
\beqa
F^{(M)}_4=d A_3.
\nonumber
\eeqa
We list all of the equations of motion and the Bianchi identities in these
supergravities in appendix A.

Let us consider a reduction of type IIA(IIB) supergravity on 
$S^{8-p}\;\;(p=0,\cdots,7)$, where $p$ takes $0,2,4,6$ for type IIA 
supergravity and $1,3,5,7$ for type IIB supergravity.
We split the ten-dimensional coordinates $X^M \;\;(M=0,1,\cdots,9)$ into
two parts, as $X^M=(\xi^\alpha,\theta_i) \;\;(\alpha=0,\cdots,p+1,\;\;
i=1,\cdots,8-p)$, where the $\xi^{\alpha}$ are $(p+2)$-dimensional coordinates
and the $\theta_i$ parametrize $S^{8-p}$. We make the following ansatz for
the ten-dimensional metric, which preserves the $(p+2)$-dimensional general
covariance:
\beqa
ds_{10} &=& G_{MN} dX^M dX^N \n
        &=& h_{\alpha\beta}(\xi)\:d\xi^{\alpha}d\xi^{\beta}
            +e^{\frac{\rho(\xi)}{2}} \: d\Omega_{8-p}.
\label{metricansatz}
\eeqa
We assume that the dilaton depends only on $\xi^{\alpha}$\footnote{It is 
sufficient for the purpose of
this paper to assume that the fields depend only on $\xi^{p+1}\:(=r)$.
However, we consider more general ansatzes such as
(\ref{metricansatz}) and (\ref{dilatonansatz}) for further developments.}:
\beqa
\phi = \phi(\xi).
\label{dilatonansatz}
\eeqa
The ansatzes for the other fields are summarized in appendix A.
Here, as an example, we perform the reduction for the $p=6$ case explicitly. 
The reductions for the other cases can be performed in the same way 
as the $p=6$ case.
The ansatzes for the other fields in the $p=6$ case are
\beqa
&&H_3=\frac{1}{3!}H_{\alpha\beta\gamma}(\xi)
d\xi^{\alpha}\wedge d\xi^{\beta} \wedge d\xi^{\gamma}
+\frac{1}{2!}d_{\alpha}(\xi)\epsilon_{\theta_{i_1}\theta_{i_2}}\:
d\xi^{\alpha}\wedge d\theta_{i_1}\wedge d\theta_{i_2}, \n 
&&F_2=\frac{1}{2!}F_{\alpha_1\alpha_2}(\xi)
d\xi^{\alpha_1}\wedge d\xi^{\alpha_2}
-\frac{1}{2!\:8!}e^{\rho/2}\varepsilon^{\alpha_1\cdots\alpha_8}
\tilde{F}_{\alpha_1\cdots\alpha_8}(\xi) 
\varepsilon_{\theta_{i_1}\theta_{i_2}}
d\theta_{i_1}\wedge d\theta_{i_2},\n
&&\tilde{F}_4=\frac{1}{4!}\tilde{F}_{\alpha_1\cdots\alpha_4}(\xi)
d\xi^{\alpha_1}\wedge\cdots\wedge d\xi^{\alpha_4} \n
&&\qquad\;\;+\frac{1}{2!\:2!\:6!}e^{\rho/2}
\varepsilon_{\alpha_1\cdots\alpha_6\beta_1\beta_2}
\tilde{F}^{\alpha_1\cdots\alpha_6}(\xi)
\varepsilon_{\theta_{i_1}\theta_{i_2}}
d\xi^{\beta_1}\wedge d\xi^{\beta_2}
\wedge d\theta_{i_1}\wedge d\theta_{i_2}.
\label{p=6ansatz}
\eeqa
By substituting (\ref{metricansatz}), (\ref{dilatonansatz}) and
(\ref{p=6ansatz}) into the equations of motion and the Bianchi identities
in type IIA supergravity in appendix A,
we obtain the following equations in eight dimensions.
\beqa
&&R_{h\alpha\beta}+2\nabla^{(8)}_{\alpha}\nabla^{(8)}_{\beta}\phi
-\frac{1}{2}\nabla^{(8)}_{\alpha}\nabla^{(8)}_{\beta}\rho
-\frac{1}{8}\pa_{\alpha}\rho\pa_{\beta}\rho
-\frac{1}{4}H_{\alpha\gamma_1\gamma_2}H_{\beta}^{\;\;\gamma_1\gamma_2}
-\frac{1}{2}e^{-\rho}d_{\alpha}d_{\beta} \n
&&-\frac{1}{2}e^{2\phi}\left(F_{\alpha\gamma}F_{\beta}^{\;\;\gamma}
+\frac{1}{3!}\tilde{F}_{\alpha\gamma_1\gamma_2\gamma_3}
             \tilde{F}_{\beta}^{\;\;\gamma_1\gamma_2\gamma_3}
+\frac{1}{5!}\tilde{F}_{\alpha\gamma_1\cdots\gamma_5}
             \tilde{F}_{\beta}^{\;\;\gamma_1\cdots\gamma_5}
+\frac{1}{7!}\tilde{F}_{\alpha\gamma_1\cdots\gamma_7}
             \tilde{F}_{\beta}^{\;\;\gamma_1\cdots\gamma_7} \right) \n
&&+\frac{1}{4}e^{2\phi}h_{\alpha\beta}(|F_2|^2+|F_4|^2+|F_6|^2+|F_8|^2)=0, \\
&&R_h+4\nabla^{(8)2}\phi-\nabla^{(8)2}\rho-4\pa_{\alpha}\phi\pa^{\alpha}\phi
-\frac{3}{8}\pa_{\alpha}\rho\pa^{\alpha}\rho+2\pa_{\alpha}\phi\pa^{\alpha}\rho
-\frac{1}{2}|H_3|^2 \n
&&-\frac{1}{2}e^{-\rho}d_{\alpha}d^{\alpha}
+e^{-\frac{\rho}{2}}R^{(S^2)}=0, \\
&&R_h+4\nabla^{(8)2}\phi-\frac{1}{2}\nabla^{(8)2}\rho
-4\pa_{\alpha}\phi\pa^{\alpha}\phi
-\frac{1}{8}\pa_{\alpha}\rho\pa^{\alpha}\rho+\pa_{\alpha}\phi\pa^{\alpha}\rho
-\frac{1}{2}|H_3|^2 
+\frac{1}{2}e^{-\rho}d_{\alpha}d^{\alpha} \n
&&-\frac{1}{2}e^{2\phi}(|F_2|^2+|F_4|^2+|F_6|^2+|F_8|^2)=0,  \\
&&\nabla^{(8)}_{\gamma}(e^{-2\phi+\frac{\rho}{2}}H^{\gamma\alpha\beta})
+\frac{1}{2}e^{\frac{\rho}{2}}\tilde{F}^{\alpha\beta\gamma_1\gamma_2}
                              F_{\gamma_1\gamma_2}
+\frac{1}{4!}e^{\frac{\rho}{2}}\tilde{F}^{\alpha\beta\gamma_1\cdots\gamma_4}
                               \tilde{F}_{\gamma_1\cdots\gamma_4}  \n           &&+\frac{1}{6!}e^{\frac{\rho}{2}}\tilde{F}^{\alpha\beta\gamma_1\cdots\gamma_6}
                               \tilde{F}_{\gamma_1\cdots\gamma_6} 
=0  \\
&&\nabla^{(8)}_{\alpha}(e^{-2\phi-\frac{\rho}{2}}d^{\alpha})
+\frac{1}{2}\varepsilon^{\alpha_1\cdots\alpha_8}\left(
+\frac{1}{6!}F_{\alpha_1\alpha_2}\tilde{F}_{\alpha_3\cdots\alpha_8} 
-\frac{1}{4!4!}\tilde{F}_{\alpha_1\cdots\alpha_4}
               \tilde{F}_{\alpha_5\cdots\alpha_8} \right)=0, \\
&&\nabla^{(8)}_{\beta}(e^{\frac{\rho}{2}}F^{\beta\alpha})
+\frac{1}{3!}e^{\frac{\rho}{2}}\tilde{F}^{\alpha\beta_1\beta_2\beta_3}
                               H_{\beta_1\beta_2\beta_3}
+\frac{1}{6!}\varepsilon^{\alpha\beta_1\cdots\beta_7} d_{\beta_1}
             \tilde{F}_{\beta_2\cdots\beta_7}=0, \\
&&\nabla^{(8)}_{\delta}(e^{\frac{\rho}{2}}\tilde{F}^{\delta\alpha\beta\gamma})
+\frac{1}{3!}e^{\frac{\rho}{2}}
\tilde{F}^{\alpha\beta\gamma\delta_1\delta_2\delta_3}
H_{\delta_1\delta_2\delta_3}
-\frac{1}{4!}\varepsilon^{\alpha\beta\gamma\delta_1\cdots\delta_5}d_{\delta_1}
             \tilde{F}_{\delta_2\cdots\delta_5}=0, \\
&&\nabla^{(8)}_{\beta}(e^{\frac{\rho}{2}}
\tilde{F}^{\beta\alpha_1\cdots\alpha_5})
+\frac{1}{3!}e^{\frac{\rho}{2}}
\tilde{F}^{\alpha_1\cdots\alpha_5\beta_1\beta_2\beta_3}
H_{\beta_1\beta_2\beta_3}
+\frac{1}{2}\varepsilon^{\alpha_1\cdots\alpha_5\beta_1\beta_2\beta_3}
d_{\beta_1}F_{\beta_1\beta_2}=0, \\
&&\nabla^{(8)}_{\beta}(e^{\frac{\rho}{2}}
\tilde{F}^{\beta\alpha_1\cdots\alpha_7})=0, \\
&&\pa_{[\alpha_1}H_{\alpha_2\alpha_3\alpha_4]}=0, \\
&&\pa_{[\alpha_1}d_{\alpha_2]}=0, \\
&&\pa_{[\alpha_1}F_{\alpha_1\alpha_2]}=0, \\
&&5\pa_{[\alpha_1}\tilde{F}_{\alpha_2\cdots\alpha_5]}
+\frac{5!}{2!\:3!}F_{[\alpha_1\alpha_2}H_{\alpha_3\alpha_4\alpha_5]}=0, \\
&&7\pa_{[\alpha_1}\tilde{F}_{\alpha_2\cdots\alpha_7]}
+\frac{7!}{3!\:4!}\tilde{F}_{[\alpha_1\cdots\alpha_4}
H_{\alpha_5\alpha_5\alpha_7]}=0.
\label{8dequations}
\eeqa
One can easily verify that these equations are derived from $I_8$, which
we will describe below. Applying this procedure to the other $p$ case, we 
obtain $I_{p+2}$. We have thus reduced type IIA(IIB) supergravity on
$S^{8-p}$ and obtain the $(p+2)$-dimensional gravities. These reductions are 
consistent truncations in the sense that every solution of $I_{p+2}$ can
be lifted to a solution of type IIA(IIB) supergravity.

Let us write down $I_{p+2}$ explicitly. We first define $\tilde{I}_{p+2}$ by
\beqa
&&\tilde{I}_{p+2}=\int d^{p+2}\xi \sqrt{-h} \left[
e^{-2\phi+\frac{8-p}{4}\rho} \left(R_h + e^{-\frac{\rho}{2}} R^{(S^{8-p})}
+4\pa_{\alpha} \phi \pa^{\alpha} \phi 
+\frac{(8-p)(7-p)}{16} \pa_{\alpha} \rho \pa^{\alpha} \rho \right.\right. \n
&&\qquad\qquad\qquad\qquad\qquad\left.\left. 
-(8-p) \pa_{\alpha} \phi \pa^{\alpha} \rho 
-\frac{1}{2} |H_3|^2 \right)
-\frac{1}{2} e^{\frac{8-p}{4}\rho} \sum_n |\tilde{F}_n|^2 \right],
\label{tildeIp+2}
\eeqa
where 
\beqa
\tilde{F}_n = F_n + s \: C_{n-3}\wedge H_3, \;\;\; 
s=\left\{ \begin{array}{rl} -1 & \mbox{for type IIA} \;\;(p=0,2,4,6) \\
                             1 & \mbox{for type IIB} \;\;(p=1,3,5,7)
          \end{array} \right.
\nonumber
\eeqa
Here $n$ takes $2,4,\cdots, p+2$ for type IIA supergravity and
$1,3,\cdots,p+2$ for type IIB supergravity. $\tilde{I}_2$ is obtained 
by setting
$H_3=0$ in (\ref{tildeIp+2}). Then, 
\beqa
&&I_{p+2}=\tilde{I}_{p+2} \;\;\; \mbox{for} \;\; p=0,1,\cdots,5 ,\\
&&I_8=\tilde{I}_8 + \int d^{8}\xi \sqrt{-h}\left(
-\frac{1}{2}e^{-\rho} |d_1|^2 \right)
+\int b_0 \left(F_2\wedge\tilde{F}_6 
-\frac{1}{2} \tilde{F}_4 \wedge \tilde{F}_4
\right),\\
&&I_9=\tilde{I}_9 + \int d^{9}\xi \sqrt{-h}\left(
-\frac{1}{2}e^{-\frac{\rho}{2}}|d_2|^2 \right)
+\int b_1 \wedge (\tilde{F}_3\wedge\tilde{F}_5
-F_1 \wedge \tilde{F}_7 ),
\eeqa
where $d_1=d b_0$ and $d_2=d b_1$.

We perform reductions of 11-d supergravity on $S^7$ and $S^4$ in
the same way as those of type IIA(IIB) supergravity. For a $S^7$ reduction,
we make the following ansatzes:
\beqa
ds_{11}&=&h_{\alpha\beta}(\xi)d\xi^{\alpha}d\xi^{\beta}
          +e^{\frac{\rho (\xi)}{2}} d\Omega_7, \n
F^{(M)}_4&=&\frac{1}{4!}F^{(M)}_{\alpha_1\cdots\alpha_4}(\xi)
             d\xi^{\alpha_1}\wedge\cdots\wedge d\xi^{\alpha_4},
\eeqa
where $\alpha$, $\beta$ run form 0 to 3.
Then, we obtain a 4-dimensional gravity, which is a consistent truncation of
11-d supergravity,
\beqa
I^{(M)}_4=\int d^4\xi \sqrt{-h} \: e^{\frac{7\rho}{4}}
\left(R_h + e^{-\frac{\rho}{2}} R^{(S^7)}
+\frac{21}{8}\pa_{\alpha}\rho\:\pa^{\alpha}\rho
-\frac{1}{2}|F^{(M)}_4|^2 \right),
\eeqa
where $F^{(M)}_4=dA_3$.
For a $S^4$ reduction, we make the following ansatzes: 
\beqa
ds_{11}&=&h_{\alpha\beta}(\xi)d\xi^{\alpha}d\xi^{\beta}
          +e^{\frac{\rho (\xi)}{2}} d\Omega_4, \n
F^{(M)}_4&=&F^{(M)}_{\alpha_1\cdots\alpha_4}(\xi)
             d\xi^{\alpha_1}\wedge\cdots\wedge d\xi^{\alpha_4} \n
          &&+\frac{1}{4!\:7!}e^{\rho} \varepsilon^{\alpha_1\cdots\alpha_7}
             \tilde{F}^{(M)}_{\alpha_1\cdots\alpha_7}(\xi)
             \varepsilon_{\theta_{i_1}\cdots\theta_{i_4}}
             d\theta_{i_1}\wedge\cdots\wedge d\theta_{i_4},
\eeqa 
where $\alpha$, $\beta$ run form 0 to 6.
Then, we obtain a 7-dimensional gravity, which is also a consistent truncation
of 11-d supergravity,           
\beqa
I^{(M)}_7=\int d^4\xi \sqrt{-h} \: e^{\rho}
\left(R_h + e^{-\frac{\rho}{2}} R^{(S^4)}
+\frac{3}{4}\pa_{\alpha}\rho\:\pa^{\alpha}\rho
-\frac{1}{2}|F^{(M)}_4|^2 -\frac{1}{2}|\tilde{F}^{(M)}_7|^2 \right),
\eeqa
where $\tilde{F}^{(M)}_7=dA_6-\frac{1}{2}A_3\wedge F^{(M)}_4$.

\vspace{1cm}

\section{Canonical formalism and the H-J equations in the reduced gravities}
\setcounter{equation}{0}
In this section, we develop the canonical formalism for $I_{p+2}$,
$I^{(M)}_4$ and $I^{(M)}_7$ obtained in the previous section.
First we rename the $(p+2)$-dimensional coordinates:
\beqa
\xi^{\mu}=x^{\mu} \;\; (\mu=0,\cdots,p), \;\;\; \xi^{p+1}=r.
\nonumber
\eeqa
Here $p$ takes 2 and 5 for $I^{(M)}_4$ and $I^{(M)}_7$, respectively.
Adopting $r$ as time, we make the ADM decomposition for the $(p+2)$-dimensional
metric.
\beqa
ds_{p+2}^2&=&h_{\alpha\beta} \: d\xi^{\alpha}d\xi^{\beta} \n
&=&(n^2+g^{\mu\nu}n_{\mu}n_{\nu})\:dr^2+2n_{\mu}\:dr\:dx^{\mu}
+g_{\mu\nu}\:dx^{\mu}dx^{\nu},
\label{ADMdecomposition}
\eeqa
where $n$ and $n_{\mu}$ are the lapse function and the shift function,
respectively. Hence force $\mu, \; \nu$ run from 0 to $p$.

In what follows, we consider a boundary surface 
specified by $r=\mbox{const.}$ and impose 
the Dirichlet condition for the fields on the boundary. Here we need to
add the Gibbons-Hawking term \cite{GH} to the actions, 
which is defined on the
boundary and ensures that the Dirichlet condition can be 
imposed consistently. 
Then, the $(p+2)$-dimensional action $I_{p+2}$ with 
the Gibbons-Hawking term on the boundary can be expressed 
in the canonical form as follows.
For $p=0,1,\cdots,5$,
\beqa
&&I_{p+2} = \int dr d^{p+1}x 
\sqrt{-g}\:\left(\pi^{\mu\nu}\pa_r g_{\mu\nu}+\pi_{\phi}\pa_r \phi
+\pi_{\rho}\pa_r \rho+\pi_B^{\mu\nu} \pa_r B_{\mu\nu} 
+\sum_n \pi_{C_{n-1}}^{\mu_1\cdots\mu_{n-1}} \pa_r C_{\mu_1\cdots\mu_{n-1}} 
\right.\n
&&\qquad \qquad \qquad \qquad \qquad  \;\;\;\left.
-nH-n_{\mu}H^{\mu}-B_{r\mu}G_B^{\mu}
-\sum_n C_{r\mu_1\cdots\mu_{n-2}}G_{C_{n-1}}^{\mu_1\cdots\mu_{n-2}}\right)
\label{cfofIp+2}
\eeqa
with
\beqa
&&H=-e^{2\phi-\frac{8-p}{4}\rho} \left( (\pi^{\mu\nu})^2
+\frac{1}{2}{\pi_{\phi}}^2+\frac{1}{2}\pi^{\mu}_{\;\; \mu}\pi_{\phi}
+\frac{4}{8-p}{\pi_{\rho}}^2+\pi_{\phi}\pi_{\rho} \right. \n
&&\qquad \left. +\left( \pi_B^{\mu\nu}
-\sum_{n\geq 3}\frac{(n-1)(n-2)}{2}C_{\mu_1\cdots\mu_{n-3}}
\pi_{C_{n-1}}^{\mu_1\cdots\mu_{n-3}\mu\nu}\right)^2 \right) \n
&&\qquad-e^{-\frac{8-p}{4}\rho} \sum_n \frac{(n-1)!}{2}
(\pi_{C_{n-1}}^{\mu_1\cdots\mu_{n-1}})^2
- {\cal L}, \n
&&H^{\mu}=-2\nabla_{\nu}\pi^{\nu\mu}+\pi_{\phi}\pa^{\mu}\phi
+\pi_{\rho}\pa^{\mu}\rho +\pi_{B}^{\nu\lambda} H^{\mu}_{\;\;\nu\lambda} \n
&&\qquad \;\;+\sum_{n\geq 4} \pi_{C_{n-1}}^{\nu_1\cdots\nu_{n-1}}
\left(F^{\mu}_{\;\;\nu_1\cdots\nu_{n-1}}
+s \frac{(n-1)!}{(n-4)!\:3!} C^{\mu}_{\;\;\nu_1\cdots\nu_{n-4}}
H_{\nu_{n-3}\nu_{n-2}\nu_{n-1}}\right),\n
&&G_B^{\mu}=-2\nabla_{\nu}\pi_B^{\nu\mu}, \n
&&G_{C_{n-1}}^{\mu_1\cdots\mu_{n-2}}=
-(n-1)\nabla_{\nu}\pi_{C_{n-1}}^{\nu\mu_1\cdots\mu_{n-2}}
-s \frac{(n+1)!}{(n-2)!\:3!} \pi_{C_{n+1}}^{\mu_1\cdots\mu_{n-2}\nu_1\nu_2\nu_3}
H_{\nu_1\nu_2\nu_3},
\label{constraints}
\eeqa
where
\beqa
&&{\cal L}=e^{-2\phi+\frac{8-p}{4}\rho} \left(R_g+4\nabla_{\mu}\nabla^{\mu}\phi
-\frac{8-p}{2}\nabla_{\mu}\nabla^{\mu}\rho-4\pa_{\mu}\phi\pa^{\mu}\phi
-\frac{(8-p)(9-p)}{16}\pa_{\mu}\rho\pa^{\mu}\rho \right.\n
&&\qquad \left.+(8-p)\pa_{\mu}\phi\pa^{\mu}\rho
-\frac{1}{2}|H_3|^2 \right)
-\frac{1}{2}e^{\frac{8-p}{4}\rho}\sum_n |\tilde{F}_n|^2
+e^{-2\phi+\frac{6-p}{4}\rho}R^{(S^{8-p})},
\label{L}
\eeqa
$I_8$ and $I_9$ with the Gibbons-Hawking terms can be rewritten in the form
(\ref{cfofIp+2}) with $p=6$ and $p=7$, respectively, up to the terms
including $b_0$ and $b_1$. The differences will turn out not to be relevant
for our purpose, so that we do not write down the precise canonical 
forms of $I_8$ and $I_9$ here.

Note that (\ref{cfofIp+2}) takes the form of (\ref{action}). Indeed, $n$, 
$n_{\mu}$ and $C_{r\mu_1\cdots\mu_{n-2}}$ play the roles of $\alpha_a$
in (\ref{action}). They give the constraints, $H=0$, $H^{\mu}=0$ and
$G_{C_{n-1}}=0$, which are called the Hamiltonian constraint, the momentum
constraint and the Gauss law constraint, respectively, and correspond to
$f_a=0$ in (\ref{eom}). Note that (\ref{cfofIp+2}) has no analogue of $H$ in
(\ref{action}). It follows from the arguments in section 2 that the 
H-J equations of this system are given by
\beqa
&&\frac{\pa S_{p+1}}{\pa \bar{r}}=0, \n
&&H=0, \n
&&H^{\mu}=0, \n
&&G_{C_{n-1}}=0
\label{Hamilton-Jacobieq}
\eeqa
with
\beqa
&&\pi^{\mu\nu}(x)=\frac{1}{\sqrt{-g(x)}}
\frac{\delta S}{\delta g_{\mu\nu}(x)},\;\;\;
\pi_{\phi}(x)=\frac{1}{\sqrt{-g(x)}}
\frac{\delta S}{\delta \phi(x)},\;\;\;
\pi_{\rho}(x)=\frac{1}{\sqrt{-g(x)}}
\frac{\delta S}{\delta \rho(x)},\n
&&\pi_{B}^{\mu\nu}(x)=\frac{1}{\sqrt{-g(x)}}
\frac{\delta S}{\delta B_{\mu\nu}(x)},\;\;\;
\pi_{C_{n-1}}^{\mu_1\cdots\mu_{n-1}}(x)=\frac{1}{\sqrt{-g(x)}}
\frac{\delta S}{\delta C_{\mu_1\cdots\mu_{n-1}}(x)},
\label{pianddelS}
\eeqa
where $\bar{r}$ is the boundary value of $r$, 
and $g_{\mu\nu}(x),\; \phi(x),\cdots,
C_{\mu_1\cdots\mu_{n-1}}(x)$ are the boundary values of the corresponding
fields. The first equation in (\ref{Hamilton-Jacobieq}) indicates that
$S_{p+1}$ does not depend on the boundary `time' explicitly.
The last three equations in (\ref{Hamilton-Jacobieq}) give functional
differential equations for $S_{p+1}$. The third and fourth ones imply
that $S_{p+1}$ must be invariant under the diffeomorphism in $p+1$ dimensions
and the $U(1)$ gauge transformations (See appendix C in Ref.\cite{ST}).
The second one is a nontrivial equation that can determine the form of
$S_{p+1}$. Hereafter, we call this equation the Hamilton-Jacobi equation.

The canonical form of $I^{(M)}_4$ with the Gibbons-Hawking term is
\beqa
I^{(M)}_4 = \int d^4 \xi \sqrt{-g} (\pi^{\mu\nu}\pa_r g_{\mu\nu}
+\pi_{\rho} \pa_r \rho +\pi_{A_3}^{\mu\nu\lambda} \pa_r A_{\mu\nu\lambda}
-n H -n_{\mu} H^{\mu} -A_{r\mu\nu} G_{A_3}^{\mu\nu} )
\label{cfofIM4}
\eeqa
with
\beqa
&&H=e^{-\frac{7}{4}\rho}\left( -(\pi^{\mu\nu})^2 
+\frac{1}{9}(\pi^{\mu}_{\;\;\mu})^2
-\frac{8}{63}\pi_{\rho}^2 +\frac{4}{9}\pi^{\mu}_{\;\;\mu}\pi_{\rho}
-3(\pi_{A_3}^{\mu\nu\lambda})^2 \right) -{\cal L}, \n
&&H^{\mu}=-2\nabla_{\nu}\pi^{\nu\mu}+\pi_{\rho}\pa^{\mu}\rho 
+\pi_{A_3}^{\nu\lambda\rho}F^{(M)\mu}_{\;\;\;\;\;\;\;\;\;\nu\lambda\rho}, \n
&&G_{A_3}^{\mu\nu}=-3\nabla_{\lambda}\pi_{A_3}^{\lambda\mu\nu},
\label{constraintsIM4}
\eeqa
where
\beqa
{\cal L}=e^{\frac{7}{4}\rho}\left(R-\frac{7}{2}\nabla_{\mu}\nabla^{\mu}\rho
-\frac{7}{2}\pa_{\mu}\rho \pa^{\mu}\rho -\frac{1}{2}|F^{(M)}_4|^2 \right)
+e^{\frac{5}{4}\rho}R^{(S^7)}.
\label{LIM4}
\eeqa
The canonical form of $I^{(M)}_7$ with the Gibbons-Hawking term is
\beqa
&&I^{(M)}_7 = \int d^7 \xi \sqrt{-g} (\pi^{\mu\nu}\pa_r g_{\mu\nu}
+\pi_{\rho} \pa_r \rho +\pi_{A_3}^{\mu\nu\lambda} \pa_r A_{\mu\nu\lambda}
+\pi_{A_6}^{\mu_1\cdots\mu_6}\pa_r A_{\mu_1\cdots\mu_6} \n
&&\qquad\qquad\qquad\qquad\;\;
-n H -n_{\mu} H^{\mu} -A_{r\mu\nu} G_{A_3}^{\mu\nu}
-A_{r\mu_1\cdots\mu_5}G_{A_6}^{\mu_1\cdots\mu_5} )
\label{cfofIM7}
\eeqa
with
\beqa
&&H=-e^{-\rho}\left( (\pi^{\mu\nu})^2-\frac{1}{9}(\pi^{\mu}_{\;\;\mu})^2
+\frac{5}{9}\pi_{\rho}^2-\frac{4}{9}\pi^{\mu}_{\;\;\mu}\pi_{\rho}
+3(\pi_{A_3}^{\mu_1\mu_2\mu_3}
+10\pi_{A_6}^{\mu_1\mu_2\mu_3\nu_1\nu_2\nu_3}A_{\nu_1\nu_2\nu_3})^2 \right.\n
&&\qquad\qquad\;\;\; \left.+\frac{6!}{2}(\pi_{A_6}^{\mu_1\cdots\mu_6})^2 
\right)-{\cal L}, \n
&&H^{\mu}=-2\nabla_{\nu}\pi^{\nu\mu}+\pi_{\rho}\pa^{\mu}\rho 
+\pi_{A_3}^{\nu\lambda\rho}F^{(M)\mu}_{\;\;\;\;\;\;\;\;\;\nu\lambda\rho}
+\left(F^{(M)\mu}_{\;\;\;\;\;\;\;\;\;\nu_1\cdots\nu_6}
-\frac{15}{2}A^{\mu}_{\;\;\nu_1\nu_2}F^{(M)}_{\nu_3\cdots\nu_6}\right)
\pi_{A_6}^{\nu_1\cdots\nu_6}, \n
&&G_{A_3}^{\mu\nu}=-3\nabla_{\lambda}\pi_{A_3}^{\lambda\mu\nu}, \n
&&G_{A_6}^{\mu_1\cdots\mu_5}=-6\nabla_{\nu}\pi_{A_6}^{\nu\mu_1\cdots\mu_5},
\label{constraintsIM7}
\eeqa
where
\beqa
{\cal L}=e^{\rho}\left(R-2\nabla_{\mu}\nabla^{\mu}\rho
-\frac{5}{4}\pa_{\mu}\rho \pa^{\mu}\rho -\frac{1}{2}|F^{(M)}_4|^2 
-\frac{1}{2}|\tilde{F}^{(M)}_7|^2 \right)
+e^{\frac{\rho}{2}}R^{(S^7)}.
\label{LIM7}
\eeqa
The H-J equations for $I^{(M)}_4$ and $I^{(M)}_7$ are derived
in the same way as the one for $I_{p+2}$.

\vspace{1cm}

\section{D$p$-brane effective action as a solution to the H-J equation}
\setcounter{equation}{0}
\subsection{D$p$-brane effective action as a solution}
In this subsection, we find a solution to the H-J equation obtained in
the previous section.
We assume that the fields are constant on the fixed-time surface. 
Let $S_{p+1}^{(0)}$ be a solution to the H-J equation under this assumption.
We can drop $b_0$ and $b_1$ consistently 
in the H-J equations for $S_7^{(0)}$ and $S_8^{(0)}$.
That is, after this simplification, 
the H-J equations for $S_7^{(0)}$ and $S_8^{(0)}$ coincide with the ones
derived from (\ref{cfofIp+2}) with $p=6$ and $p=7$, respectively.
We see from (\ref{constraints}),
(\ref{L}), (\ref{Hamilton-Jacobieq}) and (\ref{pianddelS}) that $S^{(0)}_{p+1}$
satisfies the equation
\begin{eqnarray}
&&e^{2\phi-\frac{8-p}{4}\rho} \left(
\left(\frac{1}{\sqrt{-g}}\frac{\delta S_{p+1}^{(0)}}{\delta g_{\mu\nu}}
\right)^2
+\frac{1}{2}g_{\mu\nu}\frac{1}{\sqrt{-g}}
\frac{\delta S_{p+1}^{(0)}}{\delta g_{\mu\nu}}
\frac{1}{\sqrt{-g}}\frac{\delta S_{p+1}^{(0)}}{\delta \phi}
+\frac{1}{2}\left(\frac{1}{\sqrt{-g}}
\frac{\delta S_{p+1}^{(0)}}{\delta \phi}\right)^2 \right.\n
&&\qquad\qquad\;\; +\frac{4}{8-p}\left(\frac{1}{\sqrt{-g}}
\frac{\delta S_{p+1}^{(0)}}{\delta \rho}\right)^2 
+\frac{1}{\sqrt{-g}}
\frac{\delta S_{p+1}^{(0)}}{\delta \phi}
\frac{1}{\sqrt{-g}}\frac{\delta S_{p+1}^{(0)}}{\delta \rho} \n
&&\left.\qquad\qquad\;\;
+\left(\frac{1}{\sqrt{-g}}\frac{\delta S_{p+1}^{(0)}}{\delta B_{\mu\nu}}
-\sum_n \frac{(n-1)(n-2)}{2} C_{\mu_1\cdots\mu_{n-3}}\frac{1}{\sqrt{-g}}
\frac{\delta S_{p+1}^{(0)}}{\delta C_{\mu_1\cdots\mu_{n-3}\mu\nu}}\right)^2 
\right) \n
&&+e^{-\frac{8-p}{4}\rho} \sum_n \frac{(n-1)!}{2} 
\left(\frac{1}{\sqrt{-g}}
\frac{\delta S_{p+1}^{(0)}}{\delta C_{\mu_1\cdots\mu_{n-1}}}
\right)^2 + e^{-2\phi+\frac{6-p}{4}\rho}R^{(S^{8-p})} \n
&&=0.
\label{HJ}
\end{eqnarray}

In what follows, we show that the form
\begin{equation}
S_{p+1}^{(0)}=S_{p+1}^{c}+S_{p+1}^{BI}+S_{p+1}^{WZ}+\sigma_{p+1}
\label{S0}
\end{equation}
is a solution to (\ref{HJ}), with
\begin{eqnarray}
S_{p+1}^{c}&=&\alpha_{p+1} \int d^{p+1}x \sqrt{-g} e^{-2\phi+\frac{7-p}{4}\rho}, \n
S_{p+1}^{BI}&=&\beta_{p+1} \int d^{p+1}x e^{-\phi} 
\sqrt{-\det (g_{\mu\nu}+{\cal F}_{\mu\nu})}, \n
S_{p+1}^{WZ}&=&\gamma_{p+1} \int \sum_n C_{n-1}\wedge e^{{\cal F}} \n
%&=& \gamma \int d^{p+1}x \sqrt{-g} \varepsilon^{\mu_1\cdots\mu_{p+1}}
%\left( \frac{1}{(p+1)!}C_{\mu_1\cdots\mu_{p+1}}
%+\frac{1}{2!\:(p-1)!}C_{\mu_1\cdots\mu_{p-1}}{\cal F}_{\mu_{p}\mu_{p+1}} 
%\right.\n
%&&\qquad\qquad \left.+\frac{1}{2!\:2!\:2!\:(p-3)!}C_{\mu_1\cdots\mu_{p-3}}
%{\cal F}_{\mu_{p-2}\mu_{p-1}}{\cal F}_{\mu_{p}\mu_{p+1}}+\cdots
%\right),
&=&\gamma_{p+1} \int \left(C_{p+1}+C_{p-1}\wedge{\cal F}
+\frac{1}{2}C_{p-3}\wedge {\cal F}\wedge{\cal F}+\cdots \right),
\label{ScSBISWZ}
\end{eqnarray} 
where ${\cal F}_{\mu\nu}=B_{\mu\nu}+F_{\mu\nu}^{(p+1)}$,
$F_{\mu\nu}^{(p+1)}$ is an arbitrary constant anti-symmetric tensor, 
and $\sigma_{p+1}$ 
is an  arbitrary constant. As we discussed in section 6 in Ref.\cite{ST},
$S_{p+1}^{BI}+S_{p+1}^{WZ}$ is the effective action of a probe D$p$-brane while
$S_{p+1}^c$ should be interpreted as the vacuum to vacuum amplitude and does
not contribute to the effective action of the probe D$p$-brane. Furthermore,
$F^{(p+1)}_{\mu\nu}$ is interpreted as the $U(1)$ gauge field strength
in the world-volume.

Noting that
\beqa
\frac{1}{\sqrt{-g}}\frac{\delta S_{p+1}^{(0)}}{\delta B_{\mu\nu}}
-\sum_n \frac{(n-1)(n-2)}{2} C_{\mu_1\cdots\mu_{n-3}}\frac{1}{\sqrt{-g}}
\frac{\delta S_{p+1}^{(0)}}{\delta C_{\mu_1\cdots\mu_{n-3}\mu\nu}}
=\frac{1}{\sqrt{-g}}\frac{\delta S_{p+1}^{BI}}{\delta B_{\mu\nu}} 
\nonumber
\eeqa
and 
\beqa
\frac{1}{\sqrt{-g}}\frac{\delta S_{p+1}^{WZ}}{\delta g_{\mu\nu}}=0,
\nonumber
\eeqa
one can see that the left-hand side of (\ref{HJ}) can be decomposed
into the four parts
\beq
\mbox{L.H.S. of (\ref{HJ})}=e^{2\phi-\frac{8-p}{4}\rho} \times (\;(1)+(2)+(3)\;)
+e^{-\frac{8-p}{4}\rho}\times (4) + e^{-2\phi+\frac{6-p}{4}\rho}R^{(S^{8-p})},
\label{LHS}
\eeq
with
\beqa
&&(1)=\left(\frac{1}{\sqrt{-g}}
\frac{\delta S_{p+1}^{c}}{\delta g_{\mu\nu}}\right)^2
+\frac{1}{2}g_{\mu\nu}\frac{1}{\sqrt{-g}}
\frac{\delta S_{p+1}^{c}}{\delta g_{\mu\nu}}
\frac{1}{\sqrt{-g}}\frac{\delta S_{p+1}^{c}}{\delta \phi}
+\frac{1}{2}\left(\frac{1}{\sqrt{-g}}
\frac{\delta S_{p+1}^{c}}{\delta \phi}\right)^2  \n
&&\qquad\;\;+\frac{4}{8-p}\left(\frac{1}{\sqrt{-g}}
\frac{\delta S_{p+1}^{c}}{\delta \rho}\right)^2 
+\frac{1}{\sqrt{-g}}\frac{\delta S_{p+1}^{c}}{\delta \phi}
\frac{1}{\sqrt{-g}}\frac{\delta S_{p+1}^{c}}{\delta \rho}, \n
&&(2)=2g_{\mu\lambda}g_{\nu\rho}
\frac{1}{\sqrt{-g}}\frac{\delta S_{p+1}^{c}}{\delta g_{\mu\nu}}
\frac{1}{\sqrt{-g}}\frac{\delta S_{p+1}^{BI}}{\delta g_{\lambda\rho}}
+\frac{1}{2}g_{\mu\nu}\frac{1}{\sqrt{-g}}
\frac{\delta S_{p+1}^{c}}{\delta g_{\mu\nu}}
\frac{1}{\sqrt{-g}}\frac{\delta S_{p+1}^{BI}}{\delta \phi} \n
&&\qquad \;\; +\frac{1}{2}g_{\mu\nu}
\frac{1}{\sqrt{-g}}\frac{\delta S_{p+1}^{BI}}{\delta g_{\mu\nu}}
\frac{1}{\sqrt{-g}}\frac{\delta S_{p+1}^{c}}{\delta \phi}
+\frac{1}{\sqrt{-g}}\frac{\delta S_{p+1}^{c}}{\delta \phi}
\frac{1}{\sqrt{-g}}\frac{\delta S_{p+1}^{BI}}{\delta \phi} 
+\frac{1}{\sqrt{-g}}\frac{\delta S_{p+1}^{BI}}{\delta \phi}
\frac{1}{\sqrt{-g}}\frac{\delta S_{p+1}^{c}}{\delta \rho}, \n
&&(3)=\left(\frac{1}{\sqrt{-g}}
\frac{\delta S_{p+1}^{BI}}{\delta g_{\mu\nu}}\right)^2
+\frac{1}{2}g_{\mu\nu}\frac{1}
{\sqrt{-g}}\frac{\delta S_{p+1}^{BI}}{\delta g_{\mu\nu}}
\frac{1}{\sqrt{-g}}\frac{\delta S_{p+1}^{BI}}{\delta \phi}
+\frac{1}{2}\left(\frac{1}{\sqrt{-g}}
\frac{\delta S_{p+1}^{BI}}{\delta \phi}\right)^2 \n
&&\qquad\;\;+\left(\frac{1}{\sqrt{-g}}
\frac{\delta S_{p+1}^{BI}}{\delta B_{\mu\nu}}\right)^2, \n
&&(4)=
\sum_n \frac{(n-1)!}{2} 
\left(\frac{1}{\sqrt{-g}}
\frac{\delta S_{p+1}^{WZ}}{\delta C_{\mu_1\cdots\mu_{n-1}}}\right)^2.
\eeqa
(1) and (4) are easily calculated as
\beqa
(1)&=&-\frac{7-p}{4(8-p)}\alpha_{p+1}^2 e^{-4\phi+\frac{7-p}{2}\rho}, \n
(4)&=&-\frac{1}{2}\gamma_{p+1}^2 \sum_{k=0}^{[\frac{p+1}{2}]}
\frac{(2k)!}{4^k (k!)^2}\delta^{[\mu_1}_{\nu_1} \delta^{\mu_2}_{\nu_2} 
\cdots \delta^{\mu_{2k}]}_{\nu_{2k}} {\cal F}_{\mu_1\mu_2} \cdots 
{\cal F}_{\mu_{2k-1}\mu_{2k}} 
{\cal F}^{\nu_1\nu_2} \cdots {\cal F}^{\nu_{2k-1}\nu_{2k}}.
\label{(1)and(4)}
\eeqa

In order to calculate (2) and (3), we introduce the $(p+1) \times (p+1)$
matrices ${\cal G}$
and ${\cal B}$:
\beqa
({\cal G})_{\mu\nu}=g_{\mu\nu}, \;\;\;
({\cal B})_{\mu\nu}={\cal F}_{\mu\nu}.
\nonumber
\eeqa
Then, we have
\beqa
&&\frac{1}{\sqrt{-g}}\frac{\delta S_{p+1}^{c}}{\delta g_{\mu\nu}}
=\frac{1}{2}\alpha_{p+1} e^{-2\phi+\frac{7-p}{4} \rho} \left( \frac{1}{{\cal G}} \right)^{\mu\nu},
\;\;\;
\frac{1}{\sqrt{-g}}\frac{\delta S_{p+1}^{c}}{\delta \phi}
=-2\alpha_{p+1} e^{-2\phi+\frac{7-p}{4} \rho}, \n
&&\frac{1}{\sqrt{-g}}\frac{\delta S_{p+1}^{c}}{\delta \rho}
=\frac{7-p}{4} \alpha_{p+1} e^{-2\phi+\frac{7-p}{4} \rho}, \n
&&\frac{1}{\sqrt{-g}}\frac{\delta S_{p+1}^{BI}}{\delta g_{\mu\nu}}
=\frac{1}{2}\beta_{p+1} e^{-\phi} 
\sqrt{\frac{\det ({\cal G}+{\cal B})}{\det {\cal G}}}
\left( \frac{1}{{\cal G}+{\cal B}} \: {\cal G} \: \frac{1}{{\cal G}-{\cal B}}
\right)^{\mu\nu}, \n
&&\frac{1}{\sqrt{-g}}\frac{\delta S_{p+1}^{BI}}{\delta B_{\mu\nu}}
=\frac{1}{2}\beta_{p+1} e^{-\phi} 
\sqrt{\frac{\det ({\cal G}+{\cal B})}{\det {\cal G}}}
\left( \frac{1}{{\cal G}+{\cal B}} \: {\cal B} \: \frac{1}{{\cal G}-{\cal B}}
\right)^{\mu\nu}, \n
&&\frac{1}{\sqrt{-g}}\frac{\delta S_{p+1}^{BI}}{\delta \phi}
=-\beta_{p+1} e^{-\phi} 
\sqrt{\frac{\det ({\cal G}+{\cal B})}{\det {\cal G}}}.
\eeqa
Using this notation, we can express each term in (2) and (3) in terms
of the trace of the $(p+1) \times (p+1)$ matrix and calculate 
(2) and (3) as follows: 
%\beqa
%\left(\frac{1}{\sqrt{-g}}\frac{\delta S_{p+1}^{BI}}{\delta g_{\mu\nu}}\right)^2
%=\frac{1}{4}\beta_{p+1}^2 e^{-2\phi} \frac{\det ({\cal G}+{\cal B})}{\det {\cal G}}
%\mbox{tr} \left(\frac{1}{{\cal G}+{\cal B}} \: {\cal G} \:
%\frac{1}{{\cal G}-{\cal B}} \: {\cal G} \:
%\frac{1}{{\cal G}+{\cal B}} \: {\cal G} \:
%\frac{1}{{\cal G}-{\cal B}} \: {\cal G} \: \right). 
%\eeqa
\beqa
(2)&=&\alpha_{p+1}\beta_{p+1} e^{-3\phi+\frac{7-p}{4} \rho} 
\sqrt{\frac{\det ({\cal G}+{\cal B})}{\det {\cal G}}} \left( 
\frac{1}{2}\tr \left(\frac{1}{{\cal G}} \: {\cal G} \: 
\frac{1}{{\cal G}+{\cal B}}
\: {\cal G} \: \frac{1}{{\cal G}-{\cal B}} \: {\cal G} \right) 
-\frac{p+1}{4} \right. \n
&&\qquad\qquad\qquad\qquad\qquad\qquad\qquad\;\; \left.
-\frac{1}{2}\tr \left(\frac{1}{{\cal G}+{\cal B}}
\: {\cal G} \: \frac{1}{{\cal G}-{\cal B}} \: {\cal G} \right) +2-\frac{7-p}{4} \right) \n 
&=&0, \n
(3)&=&\frac{1}{4}\beta_{p+1}^2 e^{-2\phi} 
\frac{\det ({\cal G}+{\cal B})}{\det {\cal G}} \left(
\tr \left(\frac{1}{{\cal G}+{\cal B}} \: {\cal G} \: 
\frac{1}{{\cal G}-{\cal B}} \: {\cal G} \:
\frac{1}{{\cal G}+{\cal B}} \: {\cal G} \:
\frac{1}{{\cal G}-{\cal B}} \: {\cal G} \right) \right. \n
&&\qquad\qquad\qquad\qquad\qquad\;
-\tr \left(\frac{1}{{\cal G}+{\cal B}} \: {\cal G} \: 
\frac{1}{{\cal G}-{\cal B}} \: {\cal G} \right) \n
&&\qquad\qquad\qquad\qquad\qquad\; \left.
+2 
-\tr \left(\frac{1}{{\cal G}+{\cal B}} \: {\cal B} \: 
\frac{1}{{\cal G}-{\cal B}} \: {\cal G} \:
\frac{1}{{\cal G}+{\cal B}} \: {\cal B} \:
\frac{1}{{\cal G}-{\cal B}} \: {\cal G} \right) \right) \n
&=&\frac{1}{2}\beta_{p+1}^2 e^{-2\phi} 
\frac{\det ({\cal G}+{\cal B})}{\det {\cal G}} \n
&=&\frac{1}{2}\beta_{p+1}^2 e^{-2\phi} 
\sum_{k=0}^{[\frac{p+1}{2}]} \frac{(2k)!}{4^k (k!)^2}\delta^{[\mu_1}_{\nu_1} 
\delta^{\mu_2}_{\nu_2} \cdots \delta^{\mu_{2k}]}_{\nu_{2k}} 
{\cal F}_{\mu_1\mu_2} \cdots {\cal F}_{\mu_{2k-1}\mu_{2k}} 
{\cal F}^{\nu_1\nu_2} \ldots {\cal F}^{\nu_{2k-1}\nu_{2k}}.
\label{(2)and(3)}
\eeqa
$\mbox{}$ From (\ref{HJ}), (\ref{LHS}), (\ref{(1)and(4)}) 
and (\ref{(2)and(3)}), we conclude that $S_{p+1}^{(0)}$
satisfies the H-J equation (\ref{HJ}) if 
\beq
\alpha_{p+1}^2=\frac{4(8-p)}{7-p} \: R^{(S^{8-p})}=4(8-p)^2 \;\;\; \mbox{and} \;\;\;  \beta_{p+1}^2=\gamma_{p+1}^2.
\label{condition}
\eeq

\subsection{D$p$-brane in a $B_2$ field}
In this subsection, we see that $S_{p+1}^{(0)}$ obtained in the previous
subsection reproduces the supergravity
solution representing a stack of D$p$-branes in a constant $B_2$ field.
First, we examine the cases in which $2 \leq p \leq 6$.
For simplicity, let us consider the supergravity solutions with only 
$B_{p-1 p}$ non-vanishing. These supergravity solutions were
constructed in Ref.\cite{BMM-CP}, and they are also solutions of $I_{p+2}$
taking the following forms:
\begin{eqnarray}
&&ds^2_{p+2}=
f^{-\frac{1}{2}}(\eta_{\hat{\mu}\hat{\nu}}dx^{\hat{\mu}}dx^{\hat{\nu}}
+h \delta_{ab}dx^a dx^b )+f^{\frac{1}{2}}dr^2, \n
&&e^{2\phi}=g_{st}^2 f^{-\frac{p-3}{2}}h, \;\;\;
e^{\frac{\rho}{2}}=r^2 f^{\frac{1}{2}}, \;\;\;
B_{p-1\:p}=\tan \theta f^{-1}h, \n
&&C_{01\cdots p-2}=(-1)^{p+1}g_{st}^{-1}\sin\theta f^{-1}, \;\;\;
C_{01\cdots p}=(-1)^{p+1}g_{st}^{-1}\cos\theta f^{-1} h,
\label{DpinB2}
\end{eqnarray}
where 
\begin{eqnarray}
&&\hat{\mu},\;\hat{\nu}=0,1,\cdots,p-2, \;\;\;\; a, \;b = p-1, \; p, \n
&&f=1+\frac{\tilde{Q}}{r^{7-p}}, \;\;\; 
h^{-1}=\sin^2\theta f^{-1}+\cos^2\theta.
\end{eqnarray}
Note that these supergravity solutions preserve 16 supersymmetries, and
reduce to the ordinary D$p$-brane solutions when $\theta=0$.

By varying $I_{p+2}$ with respect to the canonical momenta, we
obtain the relations between the canonical momenta and the $r$-derivatives
of the fields. Using these relations, we calculate the values of the
canonical momenta on the boundary specified by $r=\bar{r}$:
\beqa
&&-\pi_{00}=\pi_{11}= \cdots 
=\pi_{p-2 p-2}=\frac{\bar{f}^{\frac{p-5}{4}}}{g_{st}^2 \bar{h}}
\left( (8-p)\bar{r}^{7-p}\bar{f}+\frac{1}{2}\bar{r}^{8-p}\pa_{\bar{r}}\bar{f} 
\right), \n
&&\pi_{p-1 p-1}=\pi^{I_{p+2}}_{p p}
=\frac{\bar{f}^{\frac{p-5}{4}}}{g_{st}^2} 
\left( (8-p)\bar{r}^{7-p}\bar{f}+\frac{1}{2} \cos^2 \theta \bar{r}^{8-p}\bar{h} \pa_{\bar{r}} \bar{f} \right) , \n
&&\pi_{\phi}=\frac{\bar{f}^{\frac{p-3}{4}}}{g_{st}^2 \bar{h}} 
\left(4(p-8)\bar{r}^{7-p}\bar{f}-\bar{r}^{8-p} \pa_{\bar{r}} \bar{f} \right), 
\;\;\;
\pi_{\rho}=\frac{(7-p)(8-p)}{2 g_{st}^2 \bar{h}} \bar{r}^{7-p} 
\bar{f}^{\frac{p+1}{4}}, \;\;\; \pi^{I_{p+2}}_{B p-1 p}=0, \n
&&\pi_{C_{p-1}01 \cdots p-2}= \frac{(-1)^{p+1}}{(p-1)!} 
\frac{\sin\theta}{g_{st}}\bar{r}^{8-p} \bar{f}^{-\frac{p+1}{4}}
\pa_{\bar{r}} \bar{f}, \n
&&\pi_{C_{p+1}01 \cdots p}=\frac{(-1)^{p+1}}{(p+1)!} 
\frac{\cos\theta}{g_{st}}\bar{r}^{8-p} \bar{h} \bar{f}^{-\frac{p+1}{4}}
\pa_{\bar{r}} \bar{f},
\label{onshellpi1}
\eeqa
where
\beqa
\bar{f}=1+\frac{\tilde{Q}}{\bar{r}^{7-p}}, 
\;\;\; \bar{h}^{-1}=\sin^2 \theta \bar{f}^{-1} +\cos^2 \theta.
\nonumber
\eeqa

On the other hand, $S_{p+1}^{(0)}$ gives 
the following canonical momenta:
\beqa
&&\pi_{\mu\nu}=g_{\mu\lambda}g_{\nu\rho}
\frac{1}{\sqrt{-g}}\frac{\delta S_{p+1}^{(0)}}{\delta g_{\lambda\rho}} \n
&&\qquad\;\;\;
=\frac{1}{2} \alpha_{p+1} e^{-2\phi+\frac{7-p}{4}\rho} g_{\mu\nu}
+\frac{1}{2}\beta_{p+1} e^{-\phi} 
\sqrt{\frac{\det ({\cal G}+{\cal B})}{\det {\cal G}}}
\left( {\cal G} \: \frac{1}{{\cal G}+{\cal B}} \: 
{\cal G} \: \frac{1}{{\cal G}-{\cal B}} \: {\cal G}
\right)_{\mu\nu}, \n
&&\pi_{\phi}=\frac{1}{\sqrt{-g}}
\frac{\delta S_{p+1}^{(0)}}{\delta \phi}
=-2\alpha_{p+1} e^{-2\phi+\frac{7-p}{4}\rho} 
-\beta_{p+1} e^{-\phi}\sqrt{\frac{\det ({\cal G}+{\cal B})}{\det {\cal G}}}, \n
&&\pi_{\rho}=\frac{1}{\sqrt{-g}}
\frac{\delta S_{p+1}^{(0)}}{\delta \rho}
=\frac{7-p}{4}\alpha_{p+1} e^{-2\phi+\frac{7-p}{4}\rho}, \n
&&\pi_{B\mu\nu}=g_{\mu\lambda}g_{\nu\rho}
\frac{1}{\sqrt{-g}}\frac{\delta S_{p+1}^{(0)}}{\delta B_{\lambda\rho}} \n
&&\qquad \;\: = \frac{1}{2}\beta_{p+1} e^{-\phi} 
\sqrt{\frac{\det ({\cal G}+{\cal B})}{\det {\cal G}}}
\left( {\cal G} \: \frac{1}{{\cal G}+{\cal B}} \: {\cal B} \: 
\frac{1}{{\cal G}-{\cal B}} \: {\cal G}
\right)_{\mu\nu} \n
&&\qquad \;\;\;\;\; 
+\gamma_{p+1}\varepsilon_{\mu\nu \lambda_1 \lambda_2 \cdots \lambda_{p-1}} 
\sum_{k=0}^{[\frac{p+1}{2}]} 
\frac{1}{(p+1-2k)!2^k(k-1)!} C^{\lambda_1 \lambda_2 \cdots \lambda_{p+1-2k}} \n
&&\qquad\qquad\qquad\qquad\qquad\qquad\qquad
\times {\cal F}^{\lambda_{p+1-2k+1} \lambda_{p+1-2k+2}} \cdots 
{\cal F}^{\lambda_{p-2} \lambda_{p-1}}, \n
&&\pi_{C_{n-1} \mu_1 \mu_2 \ldots \mu_{n-1}}
=g_{\mu_1 \lambda_1}g_{\mu_2 \lambda_2} 
\cdots g_{\mu_{n-1} \lambda_{n-1}}
\frac{1}{\sqrt{-g}} 
\frac{\delta S_{p+1}^{(0)}}{\delta C_{\lambda_1 \lambda_2 \cdots \lambda_{n-1}}}\n
&&\qquad \qquad \qquad \;\: 
=\gamma_{p+1} 
\frac{1}{(n-1)! 2^{\frac{p-n+2}{2}} \left(\frac{p-n+2}{2}\right)!}  
\varepsilon_{\mu_1 \mu_2 \cdots \mu_{p+1}}
{\cal F}^{\mu_{n}\mu_{n+1}} 
\cdots {\cal F}^{\mu_{p}\mu_{p+1}}.
\label{pi}
\eeqa
We substitute the values of the fields in (\ref{DpinB2}) into
the right-hand sides of (\ref{pi}), setting
\beqa
F^{(p+1)}_{\mu\nu}=0.
\nonumber
\eeqa
It can be verified that
the right-hand sides of (\ref{pi}) reproduce the right-hand sides of 
(\ref{onshellpi1}) if 
\beq
\alpha_{p+1}=16-2p \;\;\; \mbox{and} \;\;\; 
\beta_{p+1}=(-1)^p \gamma_{p+1}=\frac{(p-7)\tilde{Q}\cos\theta}{g_{st}},
\label{alphabeta1}
\eeq
These conditions are consistent with (\ref{condition}).

Next, we compare the value of the on-shell action with that of $S_{p+1}^{(0)}$ 
directly.
Substituting the values of the fields with $r=\bar{r}$ in (\ref{DpinB2})
into (\ref{ScSBISWZ}), we obtain
\beqa
&&S_{p+1}^{c}=\frac{\alpha_{p+1} V_{p+1} r_0^{7-p}}{g_{st}^2}, \n
&&S_{p+1}^{BI}=\frac{\beta_{p+1} V_{p+1}}{g_{st}\cos\theta}f_0^{-1}, \n
&&S_{p+1}^{WZ}= (-1)^{p+1}\frac{\gamma_{p+1} V_{p+1}}{g_{st}\cos\theta}f_0^{-1},\label{S0value1}
\eeqa
where $V_{p+1}=\int d^{p+1}x$.
Here, we set
\beqa
\sigma_{p+1}=0.
\nonumber
\eeqa
Then, it follows from (\ref{S0}), (\ref{alphabeta1}) and
(\ref{S0value1}) that
\beqa
S_{p+1}^{(0)}=S_{p+1}^{c}+S_{p+1}^{BI}+S_{p+1}^{WZ}=\frac{(16-2p) V_{p+1} r_0^{7-p}}{g_{st}^2}
\label{S0value}
\eeqa
We calculate the values of the on-shell actions for (\ref{DpinB2})
by substituting
(\ref{onshellpi1}) with $\bar{r}$ replaced by $r$ into
(\ref{cfofIp+2}). Noting
that the constraints in (\ref{cfofIp+2}) are satisfied on shell, we 
reproduce the value of $S_{p+1}^{(0)}$ for (\ref{DpinB2}) as follows:
\beqa
I_{p+2}^{on-shell}&=&\int_0^{\bar{r}} dr d^{p+1}x \;
\sqrt{-g} (\pi^{\mu\nu}\pa_r g_{\mu\nu}+\pi_{\phi}\pa_r \phi
+\pi_{\rho}\pa_r \rho+\pi_B^{\mu\nu} \pa_r B_{\mu\nu} \n
&&\qquad \qquad \qquad  \;\;\;\;\;\;
+\sum_n \pi_{C_{n-1}}^{\mu_1\mu_{n-1}} \pa_r C_{\mu_1\cdots\mu_{n-1}} ) \n
&=&\frac{2(7-p)(8-p) V_{p+1}}{g_{st}^2} \int_0^{\bar{r}} dr \; r^{6-p} \n
&=&\frac{2(8-p) V_{p+1}}{g_{st}^2} \bar{r}^{7-p}.
\eeqa
Thus, we have shown that $S_{p+1}^{(0)}$ with
$F^{(p+1)}_{\mu\nu}=0$ and $\sigma_{p+1}=0$ reproduces 
the supergravity solution (\ref{DpinB2})
when $\alpha_{p+1}$ and $\beta_{p+1}$ take the values in
(\ref{alphabeta1}). We also verified that $S_1^{(0)}$ and $S_2^{(0)}$
reproduce the supergravity solutions representing a stack of D0-branes and
of D1-branes, respectively. Note that $S_4^{(0)}$ also reproduces the 
near-horizon limit of (\ref{DpinB2}) with $p=3$ \cite{ST}, which is
conjectured to be dual to NCYM \cite{HI,MR}.

\vspace{1cm}

\section{Effective actions of M2-brane and M5-brane as solutions to
the H-J equations}
\setcounter{equation}{0}
\subsection{M2-brane case}
We assume again that the fields are constant on the fixed-time surface.
Let us denote a solution to the H-J equation under this assumption by
$S_{M2}^{(0)}$. It follows from (\ref{cfofIM4}), (\ref{constraintsIM4})
and (\ref{LIM4}) that $S_{M2}^{(0)}$ satisfies the equation
\beqa
&&\left(\frac{1}{\sqrt{-g}}
\frac{\delta S_{M2}^{(0)}}{\delta g_{\mu\nu}}\right)^2
-\frac{1}{9}\left(g_{\mu\nu}\frac{1}{\sqrt{-g}}
\frac{\delta S_{M2}^{(0)}}{\delta g_{\mu\nu}}\right)^2
+\frac{8}{63}\left(\frac{1}{\sqrt{-g}} 
\frac{\delta S_{M2}^{(0)}}{\delta \rho}\right)^2 \n
&&\;\;\;-\frac{4}{9}g_{\mu\nu}\frac{1}{\sqrt{-g}} 
\frac{\delta S_{M2}^{(0)}}{\delta g_{\mu\nu}}
\frac{1}{\sqrt{-g}}\frac{\delta S_{M2}^{(0)}}{\delta \rho} 
+3\left(\frac{1}{\sqrt{-g}}
\frac{\delta S_{M2}^{(0)}}{\delta A_{\mu\nu\lambda}}\right)^2 
+e^{3\rho}R^{(S^7)}=0.
\eeqa
One can easily verify that the form
\beqa
S_{M2}^{(0)}=S_{M2}^{c}+S_{M2}^{NG}+S_{M2}^{WZ}+\sigma_{M2}
\eeqa
with
\beqa
&&S_{M2}^{c}=\alpha_{M2}\int d^3 x \sqrt{-g} e^{\frac{3}{2}\rho}, \n
&&S_{M2}^{NG}=\beta_{M2}\int d^3 x \sqrt{-g}, \n
&&S_{M2}^{WZ}=\gamma_{M2}\int A_3,
\eeqa
where $\sigma_{M2}$ is an arbitrary constant, satisfies the H-J equation if 
\beqa
\alpha_{M2}^2=\frac{14}{3}R^{(S^7)}=196 \;\;\; \mbox{and} \;\;\; 
\beta_{M2}^2=\gamma_{M2}^2.
\label{conditionM2}
\eeqa
Note that $S_{M2}^{NG}+S_{M2}^{WZ}$ is interpreted as
a probe M2-brane effective action as in the case of the D$p$-brane.

The supergravity solution representing a stack of M2-brane is also a solution
of $I^{(M)}_4$, which is given by
\beqa
&&ds_4^2=f^{-\frac{2}{3}}\eta_{\mu\nu}dx^{\mu}dx^{\nu}+f^{\frac{1}{3}}dr^2,
\;\;\; f=1+\frac{\tilde{Q}}{r^6}, \n
&&e^{\frac{\rho}{2}}=r^2 f^{\frac{1}{3}},\;\;\;
A_{012}=f^{-1},
\eeqa
where $\mu,\;\nu$ run from $0$ to $3$. We verified that $S_{M2}^{(0)}$
with $\sigma_{M2}=0$ reproduces this solution when
\beqa
\alpha_{M2}=14,\;\;\; \beta_{M2}=-\gamma_{M2}=-6\tilde{Q},
\eeqa
which is consistent with (\ref{conditionM2}).

\subsection{M5-brane case}
We adopt again the assumption that the fields are constant on the fixed-time
surface. Let $S_{M5}^{(0)}$ be a solution to the H-J equation under this
assumption. It follows from (\ref{cfofIM7}), (\ref{constraintsIM7}) and
(\ref{LIM7}) that $S_{M5}^{(0)}$ satisfies the equation
\beqa
&&\left(\frac{1}{\sqrt{-g}}
\frac{\delta S_{M5}^{(0)}}{\delta g_{\mu\nu}}\right)^2
-\frac{1}{9}\left(g_{\mu\nu}\frac{1}{\sqrt{-g}}
\frac{\delta S_{M5}^{(0)}}{\delta g_{\mu\nu}}\right)^2
+\frac{5}{9}\left(\frac{1}{\sqrt{-g}} 
\frac{\delta S_{M5}^{(0)}}{\delta \rho}\right)^2 \n
&&-\frac{4}{9}g_{\mu\nu}\frac{1}{\sqrt{-g}} 
\frac{\delta S_{M5}^{(0)}}{\delta g_{\mu\nu}}
\frac{1}{\sqrt{-g}}\frac{\delta S_{M5}^{(0)}}{\delta \rho} 
+3\left(\frac{1}{\sqrt{-g}}\frac{\delta S_{M5}^{(0)}}{\delta A_{\mu\nu\lambda}}
+10A_{\rho_1\rho_2\rho_3}\frac{1}{\sqrt{-g}}
\frac{\delta S_{M5}^{(0)}}{\delta A_{\mu\nu\lambda\rho_1\rho_2\rho_3}}\right)^2 \n
&&+\frac{6!}{2}
\left(\frac{\delta S_{M5}^{(0)}}{\delta A_{\mu_1\cdots\mu_6}}
\right)^2
+e^{\frac{3}{2}\rho}R^{(S^4)}=0.
\label{HJM5}
\eeqa
Let us consider the following form:
\beqa
S_{M5}^{(0)}=S_{M5}^c+S_{M5}^{BI}+S_{M5}^{WZ}+\sigma_{M5},
\label{SM5}
\eeqa
with
\beqa
&&S_{M5}^c=\alpha_{M5}\int d^6 x \sqrt{-g}e^{\frac{3}{4}\rho}, \n
&&S_{M5}^{BI}=\beta_{M5} \int d^6x \sqrt{-g}
\sqrt{1+\frac{1}{12}{\cal F}_{\mu\nu\rho}{\cal F}^{\mu\nu\rho}
+\frac{1}{288}({\cal F}_{\mu\nu\rho}{\cal F}^{\mu\nu\rho})^2
-\frac{1}{96}{\cal F}_{\mu\nu\lambda}{\cal F}^{\nu\lambda\rho}
{\cal F}_{\rho\sigma\tau}{\cal F}^{\sigma\tau\mu}}, \n
&&S_{M5}^{WZ}=\gamma_{M5}\int \left(A_6+\frac{1}{2}A_3\wedge F_3 \right),
\label{SM5s}
\eeqa
where ${\cal F}_{\mu\nu\lambda}=A_{\mu\nu\lambda}+F^{(M5)}_{\mu\nu\lambda}$,
$F^{(M5)}_{\mu\nu\lambda}$ is an arbitrary constant completely anti-symmetric
tensor, and $\sigma_{M5}$ is an arbitrary constant. 
We verified that (\ref{SM5}) satisfies (\ref{HJM5}), up to the constraint
\beqa
&&-\frac{1}{3!}\varepsilon^{\mu_1\cdots\mu_6}{\cal F}_{\mu_4\mu_5\mu_6} \n
&&=12\frac{\delta}{\delta A_{\mu_1\mu_2\mu_3}}
\sqrt{1+\frac{1}{12}{\cal F}_{\mu\nu\rho}{\cal F}^{\mu\nu\rho}
+\frac{1}{288}({\cal F}_{\mu\nu\rho}{\cal F}^{\mu\nu\rho})^2
-\frac{1}{96}{\cal F}_{\mu\nu\lambda}{\cal F}^{\nu\lambda\rho}
{\cal F}_{\rho\sigma\tau}{\cal F}^{\sigma\tau\mu}},
\label{nonlinear}
\eeqa
if 
\beqa
\alpha_{M5}^2=\frac{16}{3}R^{(S^4)}=64\;\;\; \mbox{and} \;\;\;
\beta_{M5}=-\gamma_{M5}.
\label{conditionM5}
\eeqa
The equations of motion satisfied by M5-brane are determined by the
space-time supersymmetry and the kappa symmetry \cite{HS,HS2,HSW}.
These equations of motion are equivalent to the equations derived 
from the M5-brane effective action and the non-linear self-duality 
condition \cite{W,CN}.\footnote{For a review, see Refs.\cite{SN,D}.
Also, for an alternative formulation of the M5-brane effective action, see
Refs.\cite{anotherformulation}.}
This M5-brane effective action and the non-linear self-duality condition
reduce to $S_{M5}^{BI}+S_{M5}^{WZ}$ and (\ref{nonlinear}) in our case in which
the worldvolume
of the M5-brane is 
the time-fixed surface and the static gauge is taken.

The supergravity solution of the M2-M5 bound state is given in 
Ref.\cite{ILPT-RT}, 
and it is also a solution of $I^{(M)}_7$ which takes the following form:
\beqa
&&ds_6^2=f^{-\frac{2}{3}}k^{\frac{1}{3}}\eta_{\hat{\mu}\hat{\nu}}
dx^{\hat{\mu}}dx^{\hat{\mu}} 
+f^{\frac{1}{3}}k^{-\frac{2}{3}}\delta_{ab}dx^a dx^b 
+f^{\frac{1}{3}}k^{\frac{1}{3}}dr^2, \n
&&e^{\frac{\rho}{2}}=r^2 f^{\frac{1}{3}}k^{\frac{1}{3}} ,\;\;\;
A_{012}=\sin\theta f^{-1},\;\;\; A_{345}=\tan\theta k^{-1}, \n
&&\tilde{F}^{(M)}_{012345r}=3\cos\theta \tilde{Q} r^{-4} f^{-1} k^{-1},
\eeqa
where
\beqa
&&\hat{\mu},\;\hat{\nu}=0,1,2,\;\;\; a,\;b=3,4,5, \n
&&f=1+\frac{\tilde{Q}}{r^3},\;\;\; k=\sin^2\theta+\cos^2\theta f.
\eeqa
We verified that (\ref{SM5}) with $F^{(M5)}_{\mu\nu\lambda}=0$ and
$\sigma_{M5}=0$ reproduces this solution
when $\alpha_{M5}=8$ and $\beta_{M5}=-\gamma_{M5}=-3\tilde{Q}\cos\theta$,
which is consistent with (\ref{conditionM5}).

\vspace{1cm}

\section{$(p,q)$ string and $(p,q)$ 5-brane in type IIB supergravity and
NS 5-brane in type IIA supergravity}
\setcounter{equation}{0}
\subsection{$(p,q)$ string and $(p,q)$ 5-brane}
Let us recall the $SL(2,R)$ symmetry in type IIB supergravity.
It is convenient for this purpose to work in the Einstein frame and
redefine the R-R 4-form.
The Einstein metric is given by
\beqa
G_{MN}^{(E)}=e^{-\frac{\phi}{2}}G_{MN},
\label{Einsteinmetric}
\eeqa
and the new R-R 4-form is given by
\beqa
C^{new}_4=C_4+\frac{1}{2}C_2 \wedge B_2.
\eeqa
The type IIB supergravity action (\ref{IIBaction}) is rewritten in terms of 
the Einstein metric and the new R-R 4-form:
\beqa
&&I_{IIB}=\frac{1}{2\kappa_{10}^{\:2}} \int d^{10} X \sqrt{-G^{(E)}}
\left(R_{G^{(E)}}
-\frac{1}{2}\frac{\pa_M \tau \pa^M \bar{\tau}}{(\mbox{Im}\:\tau)^2}
-\frac{1}{2}{\cal M}_{ij}\:F_3^i \: \cdot \: F_3^j -\frac{1}{4}|\tilde{F}_5|^2 
\right) \n
&& \qquad \;\;\; +\frac{\varepsilon_{ij}}{8\kappa_{10}^{\:2}} 
\int C^{new}_4 \wedge F_3^i \wedge F_3^j,
\label{IIBactionEinstein}
\eeqa
where
\beqa
&&\tau=C_0 + i \: e^{-\phi}, \n
&&{\cal M}_{ij}=\frac{1}{\mbox{Im}\:\tau} \left(
\begin{array}{cc} |\tau|^2          & \mbox{Re}\: \tau \\
             \mbox{Re}\: \tau  & 1                    
\end{array} \right), \n
&&C_2^i= \left( \begin{array}{c} B_2 \\ C_2 \end{array} \right),\;\;\;
F_3^i=d C_2^i, \n
&&\tilde{F}_5=d C^{new}_4 -\frac{1}{2} \varepsilon_{ij} C_2^i \wedge F_3^j. 
\label{defoftau}
\eeqa
One can easily check that the action (\ref{IIBactionEinstein}) and the 
self-duality condition (\ref{self-duality}) is invariant under the following
$SL(2,R)$ transformation.
\beqa
&&\tau'=\frac{a\tau+b}{c\tau+d} \;\;\;\; a,b,c,d ; \; \mbox{real},
\;\; ad-bc=1,\n
&&{C_2^i}'=\Lambda^i_{\;\;j}C_2^j, \;\;\; 
\Lambda^i_{\;\;j}=\left( \begin{array}{cc} d  & -c \\
                                           -b & a     \end{array} \right), \n
&& {C^{new}_4}'=C^{new}_4, \n
&& {G^{(E)}_{MN}}'=G^{(E)}_{MN}.
\label{SL(2,R)}
\eeqa

First, let us see how a solution to the H-J equation
corresponding to $(p,q)$ string is
obtained. Noting that (\ref{Einsteinmetric}) implies that
$\rho^{(E)}=\rho-\phi$, one can rewrite (\ref{HJ}) with $p=1$ 
in terms of the Einstein metric as follows.
\beqa
&&\left(\frac{1}{\sqrt{-g^{(E)}}}
\frac{\delta S_{2}^{(0)}}{\delta g^{(E)}_{\mu\nu}}
\right)^2
-\frac{1}{8}\left(g^{(E)}_{\mu\nu}
\frac{1}{\sqrt{-g^{(E)}}}
\frac{\delta S_{2}^{(0)}}{\delta g^{(E)}_{\mu\nu}} \right)^2
+\frac{1}{14}\left(
\frac{1}{\sqrt{-g^{(E)}}}\frac{\delta S_{2}^{(0)}}{\delta \rho^{(E)}} \right)^2
\n
&& 
-\frac{1}{2}g^{(E)}_{\mu\nu}
\frac{1}{\sqrt{-g^{(E)}}}\frac{\delta S_{2}^{(0)}}{\delta g^{(E)}_{\mu\nu}}
\frac{1}{\sqrt{-g^{(E)}}}\frac{\delta S_{2}^{(0)}}{\delta \rho^{(E)}}
+(\mbox{Im}\:\tau)^2 
\frac{1}{\sqrt{-g^{(E)}}}\frac{\delta S_{2}^{(0)}}{\delta \tau}
\frac{1}{\sqrt{-g^{(E)}}}\frac{\delta S_{2}^{(0)}}{\delta \bar{\tau}} \n
&&+g^{(E)}_{\mu\lambda}g^{(E)}_{\nu\rho}
{\cal M}^{-1}_{ij}
\frac{1}{\sqrt{-g^{(E)}}}\frac{\delta S_{2}^{(0)}}{\delta C^i_{\mu\nu}}
\frac{1}{\sqrt{-g^{(E)}}}\frac{\delta S_{2}^{(0)}}{\delta C^j_{\lambda\rho}}
+e^{3\rho^{(E)}} R^{(S^7)} \n
&&=0.
\eeqa
One can verify that this H-J equation is invariant under (\ref{SL(2,R)}),
which implies ${g^{(E)}_{\mu\nu}}'=g^{(E)}_{\mu\nu}$ and
${\rho^{(E)}}'=\rho^{(E)}$. Therefore, the $SL(2,R)$ transformed $S_2^{(0)}$
is also a solution to the H-J equation, which clearly reproduces
the supergravity solution of $(p,q)$ string.

Second, let us see briefly how a solution to the H-J equation
corresponding to $(p,q)$ 5-brane is obtained.
We consider the $p=5$ case of the reduction in section 3
with the ansatz for $H_3$ replaced by 
\beqa
&&H_3=\frac{1}{3!}H_{\alpha\beta\gamma}(\xi)
d\xi^{\alpha}\wedge d\xi^{\beta}\wedge d\xi^{\gamma} \\
&&\qquad\; +\frac{1}{3!\:7!}e^{2\phi+\frac{3}{4}\rho}
\varepsilon_{\alpha_1\cdots\alpha_7}\tilde{H}^{\alpha_1\cdots\alpha_7}(\xi)
\varepsilon_{\theta_1\theta_2\theta_3}
d\theta_{i_1}\wedge d\theta_{i_2}\wedge d\theta_{i_3}.
\eeqa
Then, we obtain as a consistent truncation a seven-dimensional gravity,
and rewrite it in terms of the Einstein metric and the new R-R 4 form:
\beqa
&&I^{(p,q) 5}_7=\int d^7\xi \sqrt{-h} e^{\frac{3}{4}\rho^{(E)}} \left(
R^{(E)}_h+e^{-\frac{1}{2}\rho^{(E)}}R^{(S^3)}
+\frac{3}{8}\pa_{\alpha}\rho^{(E)}\pa^{\alpha}\rho^{(E)}
-\frac{1}{2}\frac{\pa_{\alpha}\tau\pa^{\alpha}\bar{\tau}}{(\mbox{Im}\:\tau)^2}
\right.\n
&&\qquad\qquad\qquad\qquad\qquad\qquad\left.
-\frac{1}{2}{\cal M}_{ij}F_3^i \cdot F_3^j 
-\frac{1}{2}|\tilde{F}_5|^2
-\frac{1}{2}{\cal M}_{ij}F_7^i \cdot F_7^j \right),
\label{Ipq7}
\eeqa
where 
\beqa
&&H_3=dB_2, \;\;\; F_1=dC_0,\;\;\; F_3=dC_2,\;\;\; 
\tilde{F}_3=F_3+C_0\wedge H_3, \n
&&\tilde{F}_5=dC^{new}_4-\frac{1}{2}B_2\wedge F_3+\frac{1}{2}C_2\wedge H_3, \n
&&\tilde{F}_7=dC_6+C^{new}_4\wedge H_3-\frac{1}{2}C_2 \wedge B_2 \wedge H_3, \n
&&\tilde{H}_7=dB_6-C_0\wedge \tilde{F}_7-C^{new}_4\wedge F_3
+\frac{1}{4}C_2 \wedge C_2 \wedge H_3, \n
&& F_7^i=\left(\begin{array}{c} e^{\phi}\tilde{H}_7 \\
                                e^{-\phi}\tilde{F}_7-e^{\phi}C_0 \tilde{H}_7
                                    \end{array} \right),
\eeqa
and $\tau$, ${\cal M}_{ij}$ and $F_3^i$ are formally the same in 
(\ref{defoftau}). (\ref{Ipq7}) is invariant under the transformation
(\ref{SL(2,R)}) with 
\beqa
F_7^i=\Lambda^i_{\;\;j}F_7^j.
\label{F7}
\eeqa 
The transformation laws for $B_6$ and $C_6$ are determined by (\ref{F7}). 
Clearly, the H-J equation derived from (\ref{Ipq7}) is 
invariant under this $SL(2,R)$ transformation. 
Therefore, by rewriting (\ref{S0}) 
with $p=5$ in terms of the Einstein metric and the new R-R 4-form and applying
the $SL(2,R)$ transformation to it, we obtain a solution to the H-J equation
reproducing the supergravity solution of $(p,q)$ 5-brane.

%\beqa
%&&\left(\frac{1}{\sqrt{-g^{(E)}}}
%\frac{\delta S_{4}^{(0)}}{\delta g^{(E)}_{\mu\nu}}
%\right)^2
%-\frac{1}{8}\left(g^{(E)}_{\mu\nu}
%\frac{1}{\sqrt{-g^{(E)}}}
%\frac{\delta S_{4}^{(0)}}{\delta g^{(E)}_{\mu\nu}} \right)^2
%+\frac{3}{10}\left(
%\frac{1}{\sqrt{-g^{(E)}}}\frac{\delta S_{4}^{(0)}}{\delta \rho^{(E)}} \right)^2%\n
%&& 
%-\frac{1}{2}g^{(E)}_{\mu\nu}
%\frac{1}{\sqrt{-g^{(E)}}}\frac{\delta S_{4}^{(0)}}{\delta g^{(E)}_{\mu\nu}}
%\frac{1}{\sqrt{-g^{(E)}}}\frac{\delta S_{4}^{(0)}}{\delta \rho^{(E)}}
%+(\mbox{Im}\:\tau)^2 
%\frac{1}{\sqrt{-g^{(E)}}}\frac{\delta S_{4}^{(0)}}{\delta \tau}
%\frac{1}{\sqrt{-g^{(E)}}}\frac{\delta S_{4}^{(0)}}{\delta \bar{\tau}} \n
%&&+g^{(E)}_{\mu\lambda}g^{(E)}_{\nu\rho} \left(
%{\cal M}^{-1}_{ij}
%\frac{1}{\sqrt{-g^{(E)}}}\frac{\delta S_{4}^{(0)}}{\delta C^i_{\mu\nu}}
%\frac{1}{\sqrt{-g^{(E)}}}\frac{\delta S_{4}^{(0)}}{\delta C^j_{\lambda\rho}}
%+9{\cal M}_{ij}C^i_{\sigma\tau}C^j_{\chi\omega}
%\frac{1}{\sqrt{-g^{(E)}}}
%\frac{\delta S_{4}^{(0)}}{\delta C^{new}_{\mu\nu\sigma\tau}}
%\frac{1}{\sqrt{-g^{(E)}}}
%\frac{\delta S_{4}^{(0)}}{\delta C^{new}_{\lambda\rho\chi\omega}} \right. \n
%&&\qquad\qquad\;\;\; \left.
%+6{\cal M}_{ij} \varepsilon_{jk} C^i_{\sigma\tau}
%\frac{1}{\sqrt{-g^{(E)}}}
%\frac{\delta S_{4}^{(0)}}{\delta C^{new}_{\mu\nu\sigma\tau}}
%\frac{1}{\sqrt{-g^{(E)}}}
%\frac{\delta S_{4}^{(0)}}{\delta C^k_{\lambda\rho}} \right) \n
%&&+12\left(
%\frac{1}{\sqrt{-g^{(E)}}}
%\frac{\delta S_{4}^{(0)}}{\delta C^{new}_{\mu\nu\lambda\rho}} \right)^2 
%+e^{2\rho^{(E)}} R^{(S^5)} \n
%&&=0.
%\eeqa

\subsection{NS 5-brane in type IIA supergravity}
In this subsection, using the relation between 11-d supergravity and type IIA
supergravity, we see that the NS 5-brane effective action is a solution to
the H-J equation of supergravity and it reproduces the supergravity solution
of NS 5-brane. In order to see the relation to type IIA supergravity, we
consider a reduction of 11-d supergravity on $S^3\times S^1$, which is 
different from the one done in section 3:
\beqa
ds_{11}^2&=&G_{MN}dX^M dX^N \\
         &=&h_{\alpha\beta}(\xi)d\xi^{\alpha}d\xi^{\beta}
            +e^{\frac{1}{2}\rho_2(\xi)}d\Omega_3
            +e^{\frac{1}{2}\rho_1(\xi)}(dX^{10})^2, \n
F^{(M)}_4&=&\frac{1}{4!}F^{(M)}_{\alpha_1\cdots\alpha_4}
      d\xi^{\alpha_1}\wedge\cdots\wedge d\xi^{\alpha_4} \n
&&+\frac{1}{3!\:7!}e^{\frac{1}{4}\rho_1+\frac{3}{4}\rho_2}
\varepsilon_{\alpha_1\cdots\alpha_7}\tilde{F}^{(M)\alpha_1\cdots\alpha_7}
\varepsilon_{\theta_{i_1}\theta_{i_2}\theta_{i_3}}d\theta_{i_1}\wedge
d\theta_{i_2}\wedge d\theta_{i_3} \wedge dX^{10},
\eeqa
where $\alpha,\;\beta$ run from 0 to 6, and $X^{10}$ parametrizes $S^1$.
Then, we obtain as a consistent truncation a seven-dimensional gravity
\beqa
&&{I^{(M)}_7}'=\int d^4\xi \sqrt{-h} \: e^{\frac{1}{4}\rho_1+\frac{3}{4}\rho_2}
\left(R_h + e^{-\frac{\rho_2}{2}} R^{(S^3)}
+\frac{3}{8}\pa_{\alpha}\rho_1\:\pa^{\alpha}\rho_2
+\frac{3}{8}\pa_{\alpha}\rho_2\:\pa^{\alpha}\rho_2 \right.\n
&&\qquad\qquad\qquad\qquad\qquad\qquad\;\;\;\;\;\left.
-\frac{1}{2}|F^{(M)}_4|^2 -\frac{1}{2}|\tilde{F}^{(M)}_7|^2 \right),
\eeqa
where $\tilde{F}^{(M)}_7=dA_6-\frac{1}{2}A_3\wedge F^{(M)}_4$. 
Let us consider the form
\beqa
{S_{M5}^{(0)}}'={S_{M5}^c}'+S_{M5}^{BI}+S_{M5}^{WZ}+\sigma_{M5},
\label{SM5'}
\eeqa
with
\beqa
{S_{M5}^c}'={\alpha_{M5}}'\int d^6 x \sqrt{-g}
e^{\frac{1}{4}\rho_1+\frac{1}{2}\rho_2},
\eeqa
and $S_{M5}^{BI}$ and $S_{M5}^{WZ}$ the same in (\ref{SM5s}). One can 
verify that (\ref{SM5'}) satisfies the H-J equation of ${I^{(M)}_7}'$ under
the assumption that the fields are constant on the fixed-time surface,
up to the constraint (\ref{nonlinear}), if
\beqa
{\alpha_{M5}}'=6R^{(S^3)}=36, \;\;\;\mbox{and} \;\;\;
\beta_{M5}^2=\gamma_{M5}^2.
\eeqa

Following the relation between 11-d supergravity 
and type IIA supergravity \cite{BNS},
we define the fields in type IIA supergravity in terms of the fields in
11-d supergravity as follows.
\beqa
&&h^{(M5)}_{\alpha\beta}=e^{-\frac{2}{3}\phi}h^{(IIA)}_{\alpha\beta}, \n
&&\rho_1=\frac{8}{3}\phi, \;\;\; \rho_2=\rho-\frac{4}{3}\phi, \n
&&A_3=-C_3, \;\;\; A_6=-B_6.
\eeqa
We rewrite ${I^{(M)}_7}'$ in term of these new fields and obtain
\beqa
&&I^{(NS5)}_7=\int d^7\xi \sqrt{-h} \left[e^{-2\phi+\frac{3}{4}\rho}\left(
R_h+e^{-\frac{1}{2}\rho}R^{(S^3)}
+4\pa_{\alpha}\phi \pa^{\alpha}\phi 
+\frac{3}{8}\pa_{\alpha}\rho \pa^{\alpha}\rho
-3\pa_{\alpha}\phi \pa^{\alpha}\rho \right) \right. \n
&&\qquad\qquad\qquad\qquad\qquad\left.
-\frac{1}{2}e^{\frac{3}{4}\rho}|F_4|^2
-\frac{1}{2}e^{2\phi+\frac{3}{4}\rho}|\tilde{H}_7|^2 \right],
\eeqa
where $F_4=dC_3$ and $\tilde{H}_7=dB_6+\frac{1}{2}C_3\wedge F_4$.
This action is actually given by a consistent truncation 
of type IIA supergravity in which the ansatzes for the fields are
\beqa
&&ds_{10}^2=h_{\alpha\beta}(\xi)d\xi^{\alpha}d\xi^{\beta}
+e^{\frac{1}{2}\rho(\xi)}d\Omega_3, \n
&&\phi=\phi(\xi), \n
&&H_3=-\frac{1}{3!\:7!}e^{2\phi+\frac{3}{4}\rho}
\varepsilon_{\alpha_1\cdots\alpha_7}\tilde{H}^{\alpha_1\cdots\alpha_7}(\xi)
\varepsilon_{\theta_{i_1}\theta_{i_2}\theta_{i_3}}
d\theta_{i_1}\wedge d\theta_{i_2} \wedge d\theta_{i_3}, \n
&&F_4=\frac{1}{4!}F_{\alpha_1\cdots\alpha_4}(\xi)
d\xi^{\alpha_1}\wedge\cdots\wedge d\xi^{\alpha_4}.
\eeqa
Thus, by rewriting (\ref{SM5'}) in terms of the fields 
in type IIA supergravity, we obtain the NS 5-brane effective 
action (plus the cosmological term) that
is a solution to the H-J equation of $I^{(NS5)}_7$, a reduction of type IIA
supergravity. Clearly, this solution reproduces the supergravity solution of
NS 5-brane.

\vspace{1cm}

%\section{Conserved quantities and a search for new solutions}
%\setcounter{equation}{0}
%In this section,

%\vspace{1cm}

\section{Discussion and perspective}
\setcounter{equation}{0}
In this paper, we found that $S_{p+1}^{(0)}$ is a solution to the H-J
equation of type IIA(IIB) supergravity and it reproduces the supergravity
solution representing Dp-branes in a constant $B_2$ field. $S_{p+1}^{(0)}$
is not a complete solution to the H-J equation though it has some arbitrary
constants. In other words, it should be obtained by making some of the
arbitrary constants in a certain complete solution take specific values.
It is interesting to see
if $S_{p+1}^{(0)}$ can be generalized so that it includes more arbitrary
constants and to look for a complete solution. It is relevant to investigate
what class of supergravity solutions $S_{p+1}^{(0)}$ can reproduce.
We verified that $S_4^{(0)}$ does not reproduce the black 3-brane solution,
which preserves no supersymmetry, or the solution of
the D3-brane with the wave, which
preserves 8 supersymmetries, and that $S_6^{(0)}$ does not reproduce
the solution of the D1-D5 bound state, which preserves 8 supersymmetries.

As is clear from the general argument in section 2, 
the quantities, 
\beqa
\frac{\delta S_{p+1}^{(0)}}{\delta F^{(p+1)}_{\mu\nu}}
=\sqrt{-g}\pi_B^{\mu\nu} \;\;\; \mbox{and} \;\;\;
\frac{\delta S_{p+1}^{(0)}}{\delta \beta_{p+1}}=S_{p+1}^{BI}+S_{p+1}^{WZ},
\eeqa
are constant with respect to the time, where we take the sign in 
(\ref{condition}) such that 
$\beta_{p+1}=\gamma_{p+1}$. We verified this statement 
by an explicit calculation.
We also obtain other conserved quantities from the $SL(2,R)$
transformed $S_{p+1}^{(0)}$ in type IIB supergravity, since it includes
the continuous parameters of $SL(2,R)$. As is discussed in section 2, we can 
reduce the equations of motion to a set of the first order differential
equations by using $S_{p+1}^{(0)}$.  We may simplify these 
first order equations by using the conserved quantities so that we can
answer the above question and/or find a new solution of supergravity.
As we discussed in the introduction, it is also interesting to consider 
a reduction more complicated
than that on higher dimensional sphere and obtain another solution to
the H-J equation of supergravity, 
which should be relevant to the gauge/gravity correspondence with
less supersymmetries.

Finally, we make a remark on the case in which we perform 
a reduction on $T^{8-p}$ ($R^{8-p}$). Let us consider an ansatz for the metric
\beqa
ds_{10}^2=h_{\alpha\beta}(\xi)d\xi^{\alpha}d\xi^{\beta}
+e^{\frac{1}{2}\rho(\xi)}dy^i dy^i,
\eeqa
where $\alpha$, $\beta$ run from 0 to $p+1$, 
$i$ runs from 1 to $8-p$ and the $y^i$
parametrize $T^{8-p}$ or $R^{8-p}$. We make ansatzes for the other fields
similar to the ones in the reduction on $S^{8-p}$, and obtain
a $(p+2)$-dimensional gravity as a consistent truncation. It follows that
$(\ref{S0})$ with $\alpha_{p+1}=0$ is a solution to the H-J equation of 
this $(p+2)$-dimensional gravity. This fact seems to imply 
that the vacuum to vacuum amplitude vanishes in the reduction on the
flat manifolds.

\vspace{1cm}

\section*{Acknowledgements}
We would like to thank T. Nakatsu, M. Nishimura and C. Nunez for discussions,
and M.~Bianchi for comments on the holographic renormalization group.
A.T. is also grateful to the members of Center for Theoretical Physics, 
Massachusetts Institute
of Technology, for their hospitality, where most part of his work was done.
The work of M.S. is supported in part 
by Research Fellowships of the Japan 
Society for the Promotion of Science (JSPS) for Young Scientists (No.13-01193).

\vspace{1cm}

\section*{Appendix A: Equations of motion and Bianchi identities }
\setcounter{equation}{0}
\renewcommand{\theequation}{A.\arabic{equation}}
In this appendix, we list explicitly
the equations of motion and the Bianchi identity 
for type IIA(IIB) and 11-d supergravities. The equations of motion for
type IIA supergravity are
\beqa
&&R_{MN}+2D_M D_N \phi 
-\frac{1}{4}H_{M L_1L_2}H_N^{L_1L_2}
-\frac{1}{2}e^{2\phi}F_{ML} F^{L}_N
-\frac{1}{12}e^{2\phi}\tilde{F}_{M L_1L_2L_3}\tilde{F}_N^{L_1L_2L_3} \n
&&+\frac{1}{4}e^{2\phi}G_{MN}(|F_2|^2 +|\tilde{F}_4|^2) =0, \\
&&R +4D_M D^M \phi -4\pa_M \phi \pa^M \phi -\frac{1}{2}|H_3|^2=0, \\
&&D_L(e^{-2\phi}H^{LMN})+\frac{1}{2}F_{L_1L_2} \tilde{F}^{L_1L_2MN}) 
-\frac{1}{2!4!4!}\varepsilon^{MNL_1 \cdots L_8} 
\tilde{F}_{L_1 L_2 L_3 L_4} \tilde{F}_{L_5 L_6 L_7 L_8} =0,\\
&&D_L F^{LM} +\frac{1}{6}H_{L_1 L_2 L_3}
\tilde{F}^{M L_1 L_2 L_3}=0, \\ 
&&D_L \tilde{F}^{L M_1 M_2 M_3} 
-\frac{1}{3!4!}\varepsilon^{M_1 M_2 M_3 L_1 \cdots L_7} 
H_{L_1 L_2 L_3} \tilde{F}_{L_4 L_5 L_6 L_7} = 0,
\eeqa
where $D_M$ represents the covariant derivative in ten dimensions.
The Bianchi identities for type IIA supergravity are 
\beqa
&&dH_3=0, \\
&&dF_2=0, \\
&&d\tilde{F}_4+F_2 \wedge H_3=0.
\eeqa

The equations of motion for type IIB supergravity are
\beqa
&&R_{MN}+2D_M D_N \phi 
-\frac{1}{4}H_{M L_1L_2}H_N^{L_1L_2}
-\frac{1}{2}e^{2\phi}F_M F_N
-\frac{1}{4}e^{2\phi}\tilde{F}_{M L_1L_2}\tilde{F}_N^{L_1L_2} \n
&&-\frac{1}{4\cdot 4!}e^{2\phi}\tilde{F}_{M L_1 \cdots L_4}
\tilde{F}_N^{L_1 \cdots L_4} 
+\frac{1}{4}e^{2\phi}G_{MN}(|F_1|^2 +|\tilde{F}_3|^2)) =0, \\
&&R +4D_M D^M \phi -4\pa_M \phi \pa^M \phi -\frac{1}{2}|H_3|^2=0, \\
&&D_L(e^{-2\phi}H^{LMN})+F_L \tilde{F}^{LMN} 
+\frac{1}{6}\tilde{F}_{L_1 L_2 L_3} \tilde{F}^{MN L_1 L_2 L_3}=0, \\
&&D_L F^{L} -\frac{1}{6}H_{L_1 L_2 L_3}\tilde{F}^{L_1 L_2 L_3}=0,\\
&&D_L \tilde{F}^{LMN} -\frac{1}{6}H_{L_1 L_2 L_3}
\tilde{F}^{MN L_1 L_2 L_3}=0, \\ 
&&\tilde{F}^{M_1\cdots M_5}= 
\frac{1}{5!} \varepsilon ^{M_1\cdots M_5 L_1 \cdots L_5} 
\tilde{F}_{L_1 L_2 L_3 L_4 L_5},
\eeqa
where $D_M$ represents the covariant derivative in ten dimensions again.
The Bianchi identities for type IIB supergravity are
\beqa
&&dH_3=0, \\
&&dF_1=0, \\
&&d\tilde{F}_3-F_1 \wedge H_3=0, \\
&&d\tilde{F}_5-\tilde{F}_3 \wedge H_3=0.
\eeqa

The equations of motion for 11-d supergravity are
\beqa
&&R_{MN}-\frac{1}{12}F^{(M)}_{M L_1L_2L_3}F_N^{(M)L_1L_2L_3}
+G_{MN}\left(-\frac{1}{2}R+\frac{1}{4} |F^{(M)}_4|^2 \right)=0, \\ 
&&D_L F^{(M)L M_1 M_2 M_3} 
-\frac{1}{2 (4!)^2}\varepsilon^{M_1 M_2 M_3 L_1 \cdots L_8} 
F^{(M)}_{L_1 L_2 L_3 L_4} F^{(M)}_{L_5 L_6 L_7 L_8} = 0,
\eeqa
where $D_M$ represents the covariant derivative in eleven dimensions.
The Bianchi identities for 11-d supergravity is
\beqa
dF^{(M)}_4=0.
\eeqa

\vspace{1cm}

\section*{Appendix B: Ansatzes for the fields}
\setcounter{equation}{0}
\renewcommand{\theequation}{B.\arabic{equation}}
In this appendix, we write down the ansatzes for the fields except
the metric and the dilaton in the reduction of type IIA(IIB) supergravity
on $S^{8-p}$.

\noindent
%%%%%%%%%%%%%%%%%%%%%%%%%%%%%%%%%%%%%%%%%%%%%%%%%%%%%%%%%%%%%%%%%%%%%%
\underline{$p=0$}
\beqa
F_2=\frac{1}{2}F_{\alpha\beta}(\xi)d\xi^{\alpha}\wedge d\xi^{\beta}.
\eeqa
%%%%%%%%%%%%%%%%%%%%%%%%%%%%%%%%%%%%%%%%%%%%%%%%%%%%%%%%%%%%%%%%%%%%%%
\underline{$p=1$}
\beqa
H_3=\frac{1}{3!}H_{\alpha\beta\gamma}(\xi)
d\xi^{\alpha}\wedge d\xi^{\beta} \wedge d\xi^{\gamma},\;\;\;
F_1=F_{\alpha}(\xi)d\xi^{\alpha},\;\;\;
\tilde{F}_3=\frac{1}{3!}\tilde{F}_{\alpha\beta\gamma}(\xi)
d\xi^{\alpha}\wedge d\xi^{\beta}\wedge d\xi^{\gamma}.
\eeqa
%%%%%%%%%%%%%%%%%%%%%%%%%%%%%%%%%%%%%%%%%%%%%%%%%%%%%%%%%%%%%%%%%%%%%%
\underline{$p=2$}
\beqa
&&H_3=\frac{1}{3!}H_{\alpha\beta\gamma}(\xi)
d\xi^{\alpha}\wedge d\xi^{\beta} \wedge d\xi^{\gamma},\;\;\;
F_2=\frac{1}{2}F_{\alpha\beta}(\xi)d\xi^{\alpha}\wedge d\xi^{\beta},\n
&&\tilde{F}_4=\frac{1}{4!}\tilde{F}_{\alpha_1\cdots\alpha_4}(\xi)
d\xi^{\alpha_1}\wedge \cdots \wedge d\xi^{\alpha_4}.
\eeqa
%%%%%%%%%%%%%%%%%%%%%%%%%%%%%%%%%%%%%%%%%%%%%%%%%%%%%%%%%%%%%%%%%%%%%%%
\underline{$p=3$}
\beqa
&&H_3=\frac{1}{3!}H_{\alpha\beta\gamma}(\xi)
d\xi^{\alpha}\wedge d\xi^{\beta} \wedge d\xi^{\gamma},\;\;\;
F_1=F_{\alpha}(\xi)d\xi^{\alpha},\;\;\;
\tilde{F}_3=\frac{1}{3!}\tilde{F}_{\alpha\beta\gamma}(\xi)
d\xi^{\alpha}\wedge d\xi^{\beta}\wedge d\xi^{\gamma}, \n
&&\tilde{F}_5=\frac{1}{5!}\tilde{F}_{\alpha_1\cdots\alpha_5}(\xi)
d\xi^{\alpha_1}\wedge \cdots \wedge d\xi^{\alpha_5} \n
&&\qquad\;\;-\frac{1}{5!\:5!}e^{4\rho/5}\varepsilon^{\alpha_1\cdots\alpha_5}
\tilde{F}_{\alpha_1\cdots\alpha_5}(\xi) 
\varepsilon_{\theta_{i_1}\cdots\theta_{i_5}}
d\theta_{i_1}\wedge\cdots\wedge d\theta_{i_5}.
\eeqa
%%%%%%%%%%%%%%%%%%%%%%%%%%%%%%%%%%%%%%%%%%%%%%%%%%%%%%%%%%%%%%%%%%%%%%%%%
\underline{$p=4$}
\beqa
&&H_3=\frac{1}{3!}H_{\alpha\beta\gamma}(\xi)
d\xi^{\alpha}\wedge d\xi^{\beta} \wedge d\xi^{\gamma},\;\;\;
F_2=\frac{1}{2}F_{\alpha\beta}(\xi)d\xi^{\alpha}\wedge d\xi^{\beta},\n
&&\tilde{F}_4=\frac{1}{4!}\tilde{F}_{\alpha_1\cdots\alpha_4}(\xi)
d\xi^{\alpha_1}\wedge \cdots \wedge d\xi^{\alpha_4} \n
&&\qquad\;\;+\frac{1}{4!\:6!}e^{\rho}\varepsilon^{\alpha_1\cdots\alpha_6}
\tilde{F}_{\alpha_1\cdots\alpha_6}(\xi) 
\varepsilon_{\theta_{i_1}\cdots\theta_{i_4}}
d\theta_{i_1}\wedge\cdots\wedge d\theta_{i_4}.
\eeqa
%%%%%%%%%%%%%%%%%%%%%%%%%%%%%%%%%%%%%%%%%%%%%%%%%%%%%%%%%%%%%%%%%%%%%%%%%%
\underline{$p=5$}
\beqa
&&H_3=\frac{1}{3!}H_{\alpha\beta\gamma}(\xi)
d\xi^{\alpha}\wedge d\xi^{\beta} \wedge d\xi^{\gamma},\;\;\;
F_1=F_{\alpha}(\xi)d\xi^{\alpha},\n
&&\tilde{F}_3=\frac{1}{3!}\tilde{F}_{\alpha_1\alpha_2\alpha_3}(\xi)
d\xi^{\alpha_1}\wedge d\xi^{\alpha_2}\wedge d\xi^{\alpha_3} \n
&&\qquad\;\;+\frac{1}{3!\:7!}e^{3\rho/4}\varepsilon^{\alpha_1\cdots\alpha_7}
\tilde{F}_{\alpha_1\cdots\alpha_7}(\xi) 
\varepsilon_{\theta_{i_1}\theta_{i_2}\theta_{i_3}}
d\theta_{i_1}\wedge d\theta_{i_2} \wedge d\theta_{i_3}, \n
&&\tilde{F}_5=\frac{1}{5!}\tilde{F}_{\alpha_1\cdots\alpha_5}(\xi)
d\xi^{\alpha_1}\wedge\cdots\wedge d\xi^{\alpha_5} \n
&&\qquad\;\;-\frac{1}{2!\:3!\:5!}e^{3\rho/4}
\varepsilon_{\alpha_1\cdots\alpha_5\beta_1\beta_2}
\tilde{F}_5^{\alpha_1\cdots\alpha_5}(\xi)
\varepsilon_{\theta_{i_1}\theta_{i_2}\theta_{i_3}}
d\xi^{\beta_1}\wedge d\xi^{\beta_2}
\wedge d\theta_{i_1}\wedge d\theta_{i_2}\wedge d\theta_{i_3}.
\eeqa
%%%%%%%%%%%%%%%%%%%%%%%%%%%%%%%%%%%%%%%%%%%%%%%%%%%%%%%%%%%%%%%%%%%%%%%%%%%
\underline{$p=6$}
\beqa
&&H_3=\frac{1}{3!}H_{\alpha\beta\gamma}(\xi)
d\xi^{\alpha}\wedge d\xi^{\beta} \wedge d\xi^{\gamma}
+\frac{1}{2!}d_{\alpha}(\xi)\epsilon_{\theta_{i_1}\theta_{i_2}}\:
d\xi^{\alpha}\wedge d\theta_{i_1}\wedge d\theta_{i_2}, \n 
&&F_2=\frac{1}{2!}F_{\alpha_1\alpha_2}(\xi)
d\xi^{\alpha_1}\wedge d\xi^{\alpha_2}
-\frac{1}{2!\:8!}e^{\rho/2}\varepsilon^{\alpha_1\cdots\alpha_8}
\tilde{F}_{\alpha_1\cdots\alpha_8}(\xi) 
\varepsilon_{\theta_{i_1}\theta_{i_2}}
d\theta_{i_1}\wedge d\theta_{i_2},\n
&&\tilde{F}_4=\frac{1}{4!}\tilde{F}_{\alpha_1\cdots\alpha_4}(\xi)
d\xi^{\alpha_1}\wedge\cdots\wedge d\xi^{\alpha_4} \n
&&\qquad\;\;+\frac{1}{2!\:2!\:6!}e^{\rho/2}
\varepsilon_{\alpha_1\cdots\alpha_6\beta_1\beta_2}
\tilde{F}_4^{\alpha_1\cdots\alpha_6}(\xi)
\varepsilon_{\theta_{i_1}\theta_{i_2}}
d\xi^{\beta_1}\wedge d\xi^{\beta_2}
\wedge d\theta_{i_1}\wedge d\theta_{i_2}.
\eeqa
%%%%%%%%%%%%%%%%%%%%%%%%%%%%%%%%%%%%%%%%%%%%%%%%%%%%%%%%%%%%%%%%%%%%%%%%%%%%
\underline{$p=7$}
\beqa
&&H_3=\frac{1}{3!}H_{\alpha\beta\gamma}(\xi)
d\xi^{\alpha}\wedge d\xi^{\beta} \wedge d\xi^{\gamma}
+\frac{1}{2}d_{\alpha1\alpha2}(\xi)
d\xi^{\alpha_1}\wedge d\alpha^{\alpha_2}\wedge d\theta_1, \n
&&\tilde{F}_1=\tilde{F}_{\alpha}(\xi)
d\xi^{\alpha_1} 
-\frac{1}{9!}e^{\rho/4}\varepsilon^{\alpha_1\cdots\alpha_9}
\tilde{F}_{\alpha_1\cdots\alpha_9}(\xi) d\theta_1 \n
&&\tilde{F}_3=\frac{1}{3!}\tilde{F}_{\alpha_1\alpha_2\alpha_3}(\xi)
d\xi^{\alpha_1}\wedge d\xi^{\alpha_2}\wedge d\xi^{\alpha_3} 
+\frac{1}{2!\:7!}e^{\rho/4}
\varepsilon_{\alpha_1\cdots\alpha_7\beta_1\beta_2}
\tilde{F}_4^{\alpha_1\cdots\alpha_7}(\xi) 
d\xi^{\beta_1}\wedge d\xi^{\beta_2}\wedge d\theta_1 \n
&&\tilde{F}_5=\frac{1}{5!}\tilde{F}_{\alpha_1\cdots\alpha_5}(\xi)
d\xi^{\alpha_1}\wedge\cdots\wedge d\xi^{\alpha_5} \n
&&\qquad\;\;-\frac{1}{4!\:5!}e^{\rho/4}
\varepsilon_{\alpha_1\cdots\alpha_5\beta_1\cdots\beta_4}
\tilde{F}_5^{\alpha_1\cdots\alpha_5}(\xi)
d\xi^{\beta_1}\wedge\cdots\wedge d\xi^{\beta_4}
\wedge d\theta_1.
\eeqa
%%%%%%%%%%%%%%%%%%%%%%%%%%%%%%%%%%%%%%%%%%%%%%%%%%%%%%%%%%%%%%%%%%%%%%%%%%%%%

\end{document}